\documentclass[pra,twocolumn,superscriptaddress,a4paper]{revtex4}
\usepackage{graphicx}
\usepackage{epstopdf}
\usepackage{amssymb,amsmath}
\usepackage{epsf}
\newcommand{\bra}[1]{\left\langle #1|\right.}
\newcommand{\ket}[1]{\left.|#1\right\rangle}
\newcommand{\braket}[1]{\left\langle#1\right\rangle}

\usepackage{amssymb}
\usepackage{amsmath}

\begin{document}

\title{The dynamics and prethermalization of one dimensional quantum systems probed through 
the full distributions of quantum noise}

\author{Takuya Kitagawa}
\affiliation{Harvard-MIT Center for Ultracold Atoms, Department of Physics, Harvard University, Cambridge, Massachusetts 02138, USA}

\author{Adilet Imambekov}
\affiliation{Department of Physics and Astronomy, Rice University, Houston, Texas 77005, USA}

\author{J\"org Schmiedmayer}
\affiliation{Vienna Center for Quantum Science and Technology, Atominstitut, TU-Wien, Stadionallee 2, 1020 Vienna, Austria}

\author{Eugene Demler}
\affiliation{Harvard-MIT Center for Ultracold Atoms, Department of Physics, Harvard University, Cambridge, Massachusetts 02138, USA}

\begin{abstract}
Quantum noise correlations have been employed in several areas in physics including
condensed matter, quantum optics and ultracold atom to reveal non-classical states
of the systems. So far, such analysis mostly focused on systems in equilibrium. 
In this paper, we show that quantum noise is also a useful tool to characterize 
and study the non-equilibrium dynamics of one dimensional system. We consider the Ramsey sequence 
of one dimensional, two-component bosons, and obtain simple, analytical
expressions of time evolutions of the full distribution functions for this strongly-correlated, many-body system. 
The analysis can also be directly applied to the evolution of interference patterns between two one dimensional quasi-condensates created from a single condensate through splitting. 
Using the tools developed in this paper, we demonstrate that one dimensional dynamics in these systems 
exhibits the phenomenon known as 
"prethermalization", where the observables of {\it non-equilibrium}, long-time transient states become indistinguishable 
from those of thermal {\it equilibrium} states. 

\end{abstract}

\date{\today}

\maketitle

\section{Introduction}
Probabilistic character of Schr\"odinger wavefunctions manifests itself most directly in quantum noise. 
In many-body systems, shot-to-shot variations of experimental observables contain rich information about underlying
quantum states. Measurements of quantum noise played crucial role in establishing nonclassical states of photons
in quantum optics\cite{Aspect1982}, demonstrating quantum correlations and entanglement in electron interferometers\cite{Kindermann2007},
and verifying fractional charge of quasi-particles in quantum Hall systems\cite{Saminadayar1997,De-Picciotto1997,Dolev2008}. 
In atomic physics so far,
noise experiments focused on  systems in equilibrium. Recent work includes analysis of counting 
statistics in atom lasers\cite{Ottl2005}, establishing Hanbury-Brown-Twiss effect for both bosons and fermions\cite{Aspect2008}, analysis
of quantum states in optical lattices\cite{Hadzibabic2004, Spielman2007,Guarrera2008,Folling2005,Rom2006}, observation of momentum correlations 
in Fermi gases with pairing\cite{Greiner2005} and  investigation of thermal and quantum fluctuations 
in one and two dimensional condensates\cite{Hofferberth2008,Chabanov2001,Richard2003,Imambekov2009,Imambekov2008a,
Hadzibabic2006,Manz2010, Betz2011}. 

In this paper we demonstrate that analysis of quantum noise should also be a powerful
tool for analyzing non-equilibrium dynamics of strongly correlated systems. 
Here we study the two equivalent dynamical phenomena; one given by 
the interaction induced decoherence dynamics in Ramsey type interferometer sequences for two component Bose
mixtures in one dimension\cite{Widera2008}, and another given by the evolution of interference patterns of two one dimensional condensate created through the splitting of a single condensate\cite{Hofferberth2007, Gring2011}. We obtain 
a complete time evolution of the full distribution function of the amplitude of Ramsey
fringes or interference patterns.  In the case of Ramsey fringes, the average amplitude of Ramsey fringes
measures only the average value of the transverse spin component. On the other hand,  
full distribution functions are determined by higher order correlation functions of the spins.
Hence full distribution functions contain considerably more information about the
time evolution of the system\cite{Imambekov2007a,Polkovnikov2006,Lamacraft2008,Gring2011} and provide a powerful probe for the nature
of the quantum dynamics under study.
In particular, we use the simple expressions of full distribution functions to demonstrate the phenomena of 
"prethermalization" in these one dimensional systems, 
where observables in non-equilibrium long-time transient states become indistinguishable 
from those in thermal equilibrium states.

One dimensional systems with continuous symmetries,  including
superfluids and magnetic systems, have a special place in the family of strongly correlated systems. 
Quantum and thermal fluctuations are so extreme that long range order is not possible in equilibrium.
Such systems can not be
analyzed using standard mean-field approaches, yet they can be studied through the 
application of methods specific to one dimension such as exact
Bethe ansatz solutions\cite{Caux2006,Imambekov2008,Gangardt2003,Lieb1963,Frahm2005,Batchelor2005,IMAMBEKOV2006,Imambekov2006a,Orso2007,Tokatly2004,Fuchs2004,Lieb1963a}, effective
description using Tomonaga-Luttinger and sine-Gordon
models\cite{Iucci2010,Giamarchi2004,Cazalilla2011, Gritsev2007,DeGrandi2008,Barmettler2009}, and
numerical analysis using density-matrix renormalization group(DMRG) and matrix product state (MPS) methods\cite{Schollwock2005}.
Such systems are often considered as  general paradigms
for understanding strongly correlated systems. 
One dimensional systems also give rich examples of integrable systems, 
where due to the existence of infinite number of conserved 
quantities, equilibration does not take place\cite{Kinoshita2006,Rigol2008,Cassidy2009}. 
Hence the problem we consider in this paper is important for 
understanding fundamental issues such as the quantum dynamics of strongly correlated systems 
and equilibration/non-equilibration of many-body systems, as well as for 
possible applications of spinor condensates in
spectroscopy, interferometry, and quantum information processing
\cite{Wineland1994, Sørensen2001, Kitagawa2010}.

Our work is motivated by recent experiments of Widera {\it et
al.}\cite{Widera2008}  who used two hyperfine states of $^{87}$Rb atoms
confined in 2D arrays of one dimensional tubes 
to perform Ramsey type interferometer sequences. They
observed rapid decoherence of Ramsey
fringes and the near absence of spin echo. Their results could not be explained
within the single mode approximation which assumes a macroscopic
Bose condensation into a single orbital state, but could be
understood in terms of the multi-mode Tomonaga-Luttinger type model. Yet the
enhanced decoherence rate and suppression of spin echo do not
provide unambiguous evidences for what the origin of decoherence is. In
this paper, we suggest that the crucial evidence of the multi-mode
dynamics as a source of decoherence should come from the time
evolution of the full distribution functions  of the Ramsey
fringe amplitude.  Such distribution functions should be
accessible in experiments on Atom Chips\cite{Folman2000,Reichel2011,Hofferberth2008,Bederson2002,Wicke2010} from
the analysis of shot-to-shot fluctuations. 

This paper is organized as follows. In Section \ref{section:qualitative}, 
we describe two physically distinct, but yet mathematically equivalent, dynamics in
one dimensional systems, namely, the dynamics of spins in a Ramsey sequence and
the dynamics of phase and contrast in interference patterns between
two split condensates created from a single condensate. We start with illustrating the basic physics governing the dynamics studied in this paper, and give a summary of the central results including the prediction of prethermalization phenomena. The formal descriptions of the details of the theory for spin dynamics in Ramsey sequence start from Section \ref{section:hamiltonian},
where we give the Hamiltonian of the one dimensional system based on Tomonaga-Luttinger approach. 
In Section \ref{section:nomix}, we derive the analytical expression for the time evolution of the
full distribution function for a simple case in which charge and spin degrees of 
freedom decouple. This decoupling limit gives a good approximation to the experimental situation of Widera
{\it et al}\cite{Widera2008}. A short summary of the result in this decoupling limit 
has been already reported in Ref \cite{Kitagawa2010}.
More general case in which spin and charge degrees of freedom mix is studied in Section \ref{section:mix}. Such mixing introduces the dependence of spin distribution functions
on the initial temperature of the system.
All the results obtained in Sections \ref{section:nomix} and \ref{section:mix} can be extended to
the study of the dynamics of interference patterns, using the mapping described in Section \ref{section:qualitative}. 
The details of the dynamics of phases and contrasts in interference patterns between split condensates is 
studied in Section \ref{section:interference}. We demonstrate the prethermalization phenomena and 
show that the interference contrasts of split condensates in a steady state have indistinguishable 
distributions from those of thermal condensates at some effective temperature $T_{\textrm{eff}}$. 
We conclude in Section \ref{section:conclusion} with a discussion of possible extensions of this work.

\section{Description of one dimensional dynamics and the summary of results} \label{section:qualitative}

\subsection{Ramsey Dynamics} 

\begin{figure}[t]
\begin{center}
\includegraphics[width = 7cm]{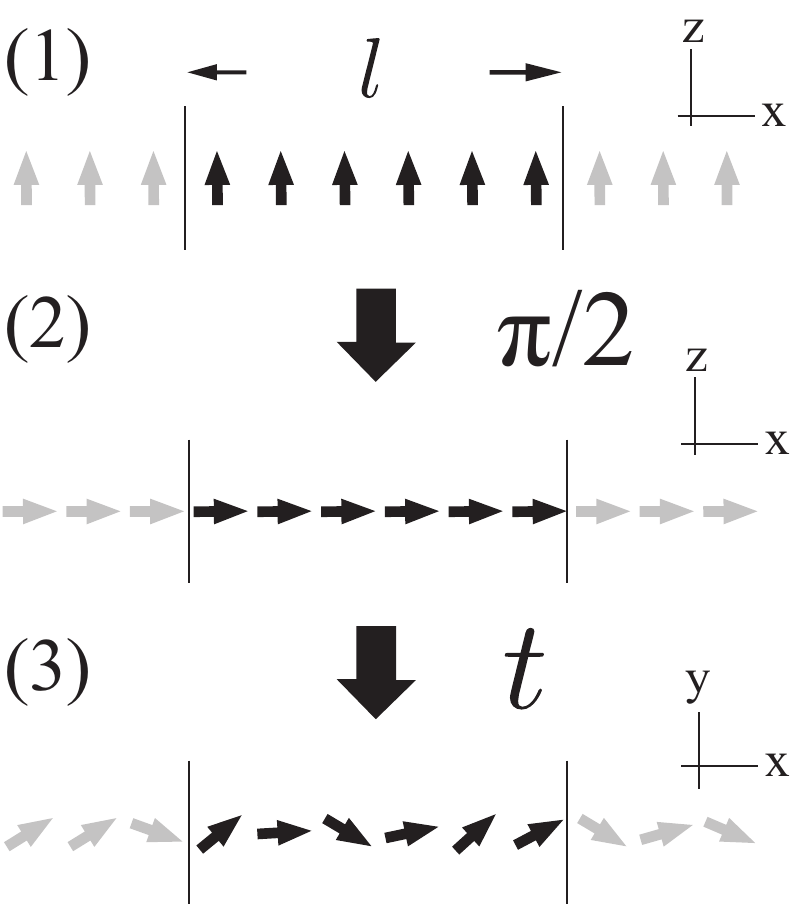}
\caption{Ramsey sequence 
for one dimensional system with two component 
bosons considered in this paper. 
(1) All atoms are
prepared in spin up state; (2) $\pi/2$ pulse is applied to
rotate each atom into the $x$ direction; (3) spins freely evolve
for time $t$. In actual experiments, final $\pi/2$ pulse is applied 
to measure the $x$ component of spin. The imaging step (4) 
is omitted in the illustration. In this paper, spin operators
refer to the ones before the final $\pi/2$ pulse.}
\label{ramsey}
\end{center}
\end{figure}

In this paper, we study the dynamics of one dimensional, interacting two-component Bose mixtures 
in the Ramsey-type sequence. In analogy with spin-$1/2$ particles, we refer
to one component to be spin-up and the other component to be spin-down. 
In the experiment of cold atoms in Ref \cite{Widera2008}, 
two hyperfine states are used for these two components. 
In the following, we consider a generic situation where there is no symmetry that 
relates spin-up and spin-down. In particular, unlike fermions with spin-$1/2$, 
there is no SU$(2)$ symmetry. In a typical experimental setup with cold atoms, 
there is a harmonic confinement potential along the longitudinal direction of 
condensates, but here we assume the absence of such a 
harmonic trap potential. Our consideration gives 
a good approximation for the central region of cold atom experiments in the presence of 
such potentials. 

Ramsey-type sequence is described as follows(Figure \ref{ramsey}): 
\begin{enumerate}
\item All atoms are prepared in spin up state at low temperature
\item $\pi/2$ pulse is applied to rotate the spin of each atom into the $x$ direction
\item Spins evolve for time $t$
\item Spins in the transverse direction ($x-y$ plane) are measured
\end{enumerate}
In a typical experimental situations\cite{Widera2008}, the last measurement step 
is done by applying a $\pi/2$ pulse to map the transverse spin component 
into $z$ direction, which then can be measured. In the following discussions, 
we describe the dynamics in the rotating frame of Larmor frequency in which
the chemical potentials of spin-up and spin-down are the same in the absence of 
interactions. In this frame, the evolution of spins in the third step is dictated by
the diffusion dynamics coming from interactions. Unlike the conventional use of the Ramsey sequence
in the context of precision measurements, here we employ the Ramsey sequence as a probe 
of correlation functions in one dimensional system. 

The description of the spin dynamics starts from the highly excited state prepared after the 
$\pi/2$ pulse of step 2. The subsequent dynamics during step 3 crucially depends on 
the nature of excitations in the system. In particular, the dynamics of two-component
Bose mixture in one dimension is quite different from those in three dimension. 
In three dimensions, bosons form a Bose-Einstein condensate(BEC) at low temperature, 
and particles occupy a macroscopic number of $k=0$ mode. Then, the spin diffusion of three dimensional BEC
is dominated by the {\it spatially homogeneous} dynamics coming from the single $k=0$ mode at sufficiently
low temperatures. On the other hand, bosonic systems in one dimension do not have the macroscopic
occupancy of the $k=0$ mode, and their physics is dominated by the strong fluctuations, to the extent that 
the system cannot retain the long range phase coherence even at zero temperature\cite{Giamarchi2004}.
Thus, the spin dynamics of one dimensional bosonic system necessarily involves a large number of 
modes with different momenta and the spin becomes {\it spatially inhomogeneous} during the step 3 above.

Such dynamics unique to one dimension can be probed through the
observation of transverse spin components in the fourth step. Since we aim to capture the multi-mode nature 
of the dynamics in one dimension, we consider the observation of spins at length scale $l$,
given by 
\begin{eqnarray}
\hat{S}^a_{l}(t) = \int_{-l/2}^{l/2} dr \hat{S}^a(r, t)
\end{eqnarray}
where $\hat{S}^a(r, t)$ with $a=x, y$ are the transverse components of spin operators after time 
evolution of step $3$ of duration $t$. We assume that $l$ is much larger than the spin healing length $\xi_{s}$,
and much smaller than the system size $L$ to avoid finite size effects. Furthermore, we assume that
the number of particles within the length $l$, $N_l$, is large, so that the simultaneous measurements of 
$\hat{S}^x_{l}(t)$ and $\hat{S}^y_{l}(t)$ are in principle possible. For large $N_l$, the non-commutativity of 
$\hat{S}^x_{l}$ and $\hat{S}^y_{l}$ gives corrections of the order of $1/\sqrt{N_l}$ compared to the average values. 
In this situation, it is also possible to measure the magnitude of transverse spin components, 
$\hat{S}^{\perp}_{l} = \sqrt{ \left( \hat{S}^x_{l} \right)^2 + \left( \hat{S}^y_{l} \right)^2}$, which we will extensively study
in the later sections. 

Due to quantum and thermal fluctuations, the measurements of $\hat{S}^a_{l}(t)$ give different values from shot-to-shot.
After the $\pi/2$ pulse of step 2, the spins are prepared in $x$ direction, so the average value yields
$\braket{\hat{S}^x_{l}(t=0)} \approx N_{l}/2$ and $\braket{\hat{S}^y_{l}(t=0)} \approx 0$. 
In the rotating frame of Larmor frequency,
the subsequent evolution does not change the expectation value of the $y$ component 
so that $\braket{\hat{S}^y_{l}(t)} \sim 0$ throughout. 
The decay of 
the average $\braket{\hat{S}^x_{l}(t)}$ during the evolution in step 3 tells us the strength
of spin diffusion in the system. The behaviors of $\braket{\hat{S}^a_{l}(t)}$ due to spin diffusion are similar 
for one and three dimensions, and the difference is quantitative,
rather than qualitative. 
On the other hand, a richer information about the
dynamics of one dimensional system is contained in the noise of $\hat{S}^a_{l}(t)$. 
Such noise inherent to quantum systems is captured by higher moments $\langle \, (\, \hat{S}^a_{l}(t) \,)^n \rangle$.
In this paper, we obtain the expression for the full distribution function $P_{l}^{a}(\alpha,t)$ which 
can produce any moments of $\hat{S}^a_{l}(t)$ through the relation 
\begin{eqnarray}
\langle \, (\, \hat{S}^a_{l}(t) \,)^n \rangle = \int \, d\alpha \, P^{a}_{l}(\alpha,t)  \alpha^n,
\label{moments}
\end{eqnarray}
where $P_{l}^{a}(\alpha,t) d\alpha $ represents the probability that the measurement of 
$\hat{S}^a_{l}(t)$ gives the value between $\alpha$ and $\alpha +d\alpha$. 
We will see in Section \ref{section:nomix} that 
it is also possible to obtain the joint distributions $P_{l}^{x,y}(\alpha,\beta, t)$ 
of $\hat{S}^x_{l}(t)$ and $\hat{S}^y_{l}(t)$ as well as the distribution $P_{l}^{\perp}(\alpha, t)$ 
of the squared transverse magnitude $\left( \hat{S}^{\perp}_{l} \right)^2$. 

Now we summarize the main results of this paper, and 
give a qualitative description of spin dynamics in the Ramsey sequence.
Elementary excitations of spin modes in the system are described in terms of 
linearly dispersing spin waves with momenta $k$ and excitation energies $c_{s} |k|$,
where $c_{s}$ is the spin wave velocity.  
When certain symmetry conditions are satisfied(see discussion in Section \ref{section:nomix}), spin
and charge degrees of freedom decouple, and these 
spin waves are free and they do not interact among themselves 
in the low energy descriptions within so-called Tomonaga-Luttinger theory\cite{Giamarchi2004,Cazalilla2011}. 
Here, we describe the result in this decoupling limit, but the qualitative picture
does not change even after the coupling between spin and charge is introduced, as we will see in Sec.~\ref{section:mix}.

\begin{figure}[th]
\begin{center}
\includegraphics[width = 8cm]{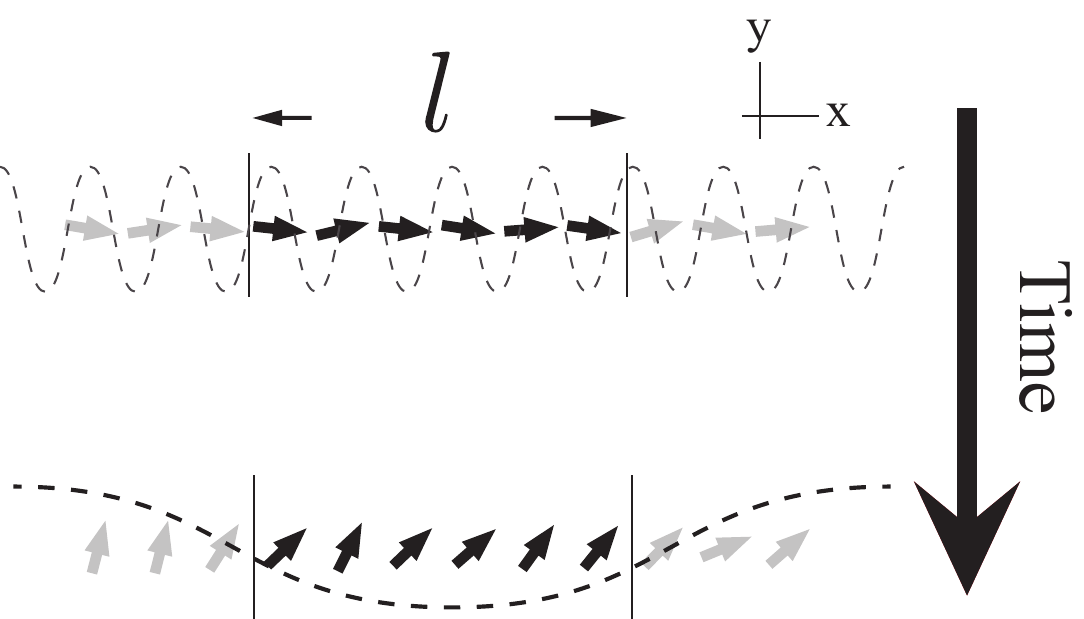}
\caption{Illustration of the dynamics of spins in the presence of spin wave excitations. 
At short times(top), high momenta excitations contribute to fluctuations of the spins,
but their effect is weak. At long times(bottom), low momenta excitations lead to the strong fluctuations of the spins. 
Such fluctuations with wavelengths larger than $l$ rotate the regions of length $l$ 
as a whole so that they do not lead to the decay of the magnitude of spin $\hat{S}^{\perp}_{l}$,
but result in diffusion of $\hat{S}^{x}_{l}$. }
\label{illustration}
\end{center}
\end{figure}

\begin{figure}[t]
\begin{center}
\includegraphics[width = 8.5cm]{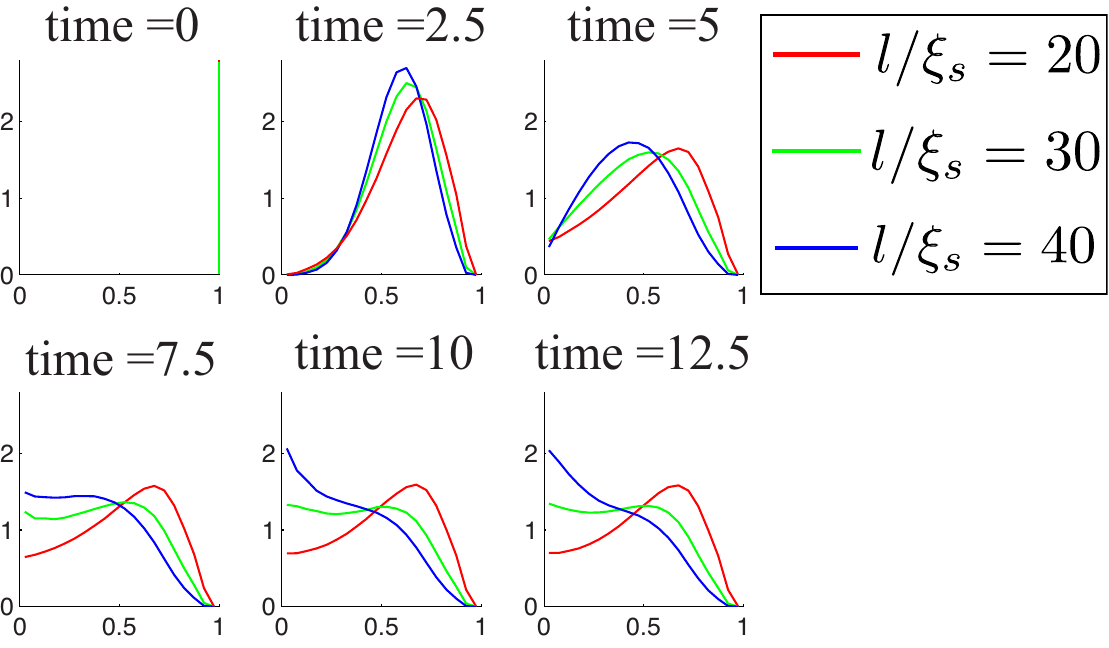}
\caption{ Time evolution of the distribution $P_{l}^{\perp}(\alpha)$
of the squared transverse magnitude of spins,$\left( \hat{S}^{\perp}_{l} \right)^2$,
for the system size $L/\xi_{s} =200$, the spin Luttinger parameter $K_{s} =20$ and 
various integration length $l/\xi_{s} =20,30,40$. Here $\xi_{s}$ is the spin healing length, 
and the $x$ axis is scaled such that the maximum
value of $\alpha$ is $1$. Time is measured in units of 
$\xi_{s}/c_{s}$ where $c_{s}$ is the spin sound wave velocity. The evolution of the distribution
crucially depends on the integration length. The steady state of the distribution of 
the squared transverse magnitude has a peak at 
a finite value for short integration length $l/\xi_{s} =20$, 
whereas the peak is at $0$ for long integration length $l/\xi_{s} =40$. }
\label{fig:figure1}
\end{center}
\end{figure}

\begin{figure*}[t]
\begin{center}
\includegraphics[width = 17cm]{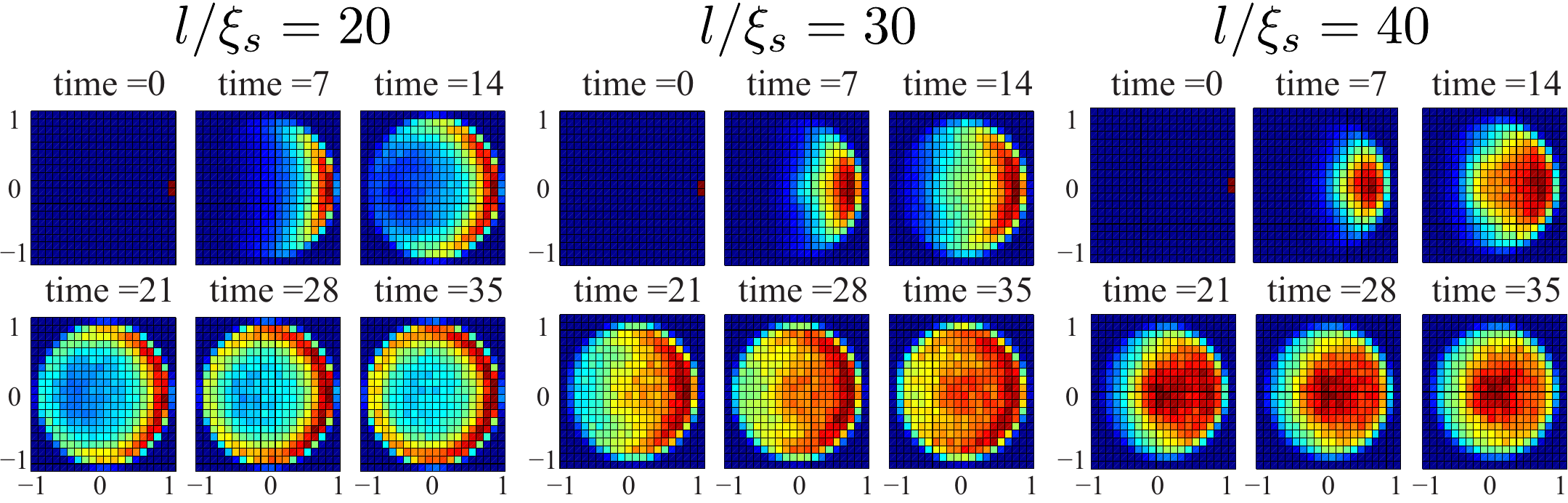}
\caption{ Time evolution of the joint distribution function $P^{x,y}(\alpha,\beta)$
for the system size $L/\xi_{s} =200$, the spin Luttinger parameter $K_{s} =20$ and 
various integration lengths $l/\xi_{s} =20$(left),$30$(middle), and $40$(right). Time is measured in units of 
$\xi_{s}/c_{s}$ where $c_{s}$ is the spin sound wave velocity. Here axes are scaled such that the maximum
value of $\alpha$ and $\beta$ are $1$.
For for short integration length $l/\xi_{s} =20$, the dynamics leads to the distribution with the "ring"-like 
structure, showing that the magnitude of spins does not decay much (spin diffusion regime). 
On the other hand, for longer integration lengths, the magnitude of spins decays quickly
and the distribution forms a "disk"-like structure(spin decay regime). }
\label{fig:jointdist}
\end{center}
\end{figure*}

The initial state prepared after $\pi/2$ pulse in step 2 
in which all spins point in the $x$ direction is far from the equilibrium state of the system
because interactions of spins are not symmetric in terms of spin rotations. 
Thus, the initial state contains many excitations 
and the subsequent dynamics of spins is determined by time evolution of the spin waves. 
A spin wave excitation with momentum $k$ rotates spins with length scale $\sim 2\pi/k$ 
and time scale $\sim 1/(c_{s} |k|)$. 
The amplitude of fluctuations coming from the spin wave with momentum $k$ is 
determined by the initial state as well as the nature of spin wave excitations. 
We find that the energy stored in each mode is approximately the same(see discussion in Sec.~\ref{section:interpretation}), 
thus the amplitude of fluctuations for wave vector $k$ scales as $1/k^2$. Therefore, the fluctuation
of spins is weak at short wave lengths and short times, and strong at long wave lengths and
long times. In Fig.\ref{illustration}, we illustrate such dynamics of spins due to fluctuations of spin wave
excitations. 
It leads to the distributions presented in Fig.\ref{fig:figure1} and  Fig.\ref{fig:jointdist}. 
Here, we have plotted distribution function of the squared transverse magnitude of spins
$\left( \hat{S}^{\perp}_{l} \right)^2$(Fig.\ref{fig:figure1}) and the joint distribution function(Fig.\ref{fig:jointdist})
with $L=200$, $K_{s} =20$ and various integration length $l/\xi_{s} =20,30,40$. 
$K_{s}$ is the spin Luttinger parameter, which measures the strength of correlations in 1D
system (see Eq. (\ref{spinhamiltonian}) below), and $\xi_{s}$ is a spin healing length which gives a characteristic
length scale in the low energy theory of spin physics.

The multi-mode nature of one dimensional system, in which spin correlations
at different length scales are destroyed in qualitatively different fashion during the dynamics,
can be revealed most clearly in the squared transverse magnitude of 
spins $\left( \hat{S}^{\perp}_{l} \right)^2$, plotted in Fig.\ref{fig:figure1}. 
In the initial state, all the spins are aligned in the $x$ direction, so the distribution 
of $\left( \hat{S}^{\perp}_{l} \right)^2$ is a delta function peak at its maximum value, $\sim (\rho l)^2$,
where $\rho$ is the average density of spin-up or spin-down. The evolution of spin waves
 lead to the fluctuations of spins and thus to the decay of the integrated magnitude of the transverse spin. 
 How the spin waves affect the integrated magnitude of spins 
strongly depends on the wavelength of the excitations.
Spin excitations with momenta much smaller than 
$\sim 2\pi/l$ do not affect $\left( \hat{S}^{\perp}_{l} \right)^2$ since these spin waves rotate the spins 
within $l$ as a whole, while spin excitations with higher momenta lead to the decay
of the magnitude. This is in stark contrast with the $x$ component of the spin $\hat{S}^{x}_{l}$, which receives 
contributions from spin waves of all wavelengths.

As a result of different contributions of spin wave excitations with different wavelengths 
to the integrated spin magnitude,
there are two distinct behaviors of the distributions of $\left( \hat{S}^{\perp}_{l} \right)^2$;
one for short integration length $l$, which we call "spin diffusion regime" and 
another for long integration length $l$, which we call "spin decay regime." 

For short integration length $l$, the distribution function of $\left( \hat{S}^{\perp}_{l} \right)^2$
is always peaked near its maximum value $(\rho l)^2$ during the dynamics
because the strengths of fluctuations coming from spin waves with high momenta 
are suppressed by $1/k^2$(Fig.\ref{fig:figure1}, $l/\xi_{s}=20$). 
While the magnitude of spins does not decay in this regime, fluctuations still lead to a diffusion
of $\hat{S}^{x}_{l} $, thus, we call this regime the "spin diffusion regime."

On the other hand, for long integration length, spin waves lead to fluctuations of the spins within the integration region,
and the spins are 
randomized after a long time. This randomization of spins leads to the development of 
a Gaussian-like peak near $\left( S^{\perp}_{l} \right)^2 =0$(Fig.\ref{fig:figure1}, $l/\xi_{s}=40$). 
During the intermediate time, 
both peaks at $0$ and the maximum value $(\rho l)^2$ are present, and one can observe
the double peak structure. Because of the strong decaying behavior of the magnitude of spin,
we call this regime the "spin decay regime."

More complete behaviors of distribution functions can be captured by looking at the 
joint distribution functions from which we can read off the distributions of both 
$\left( \hat{S}^{\perp}_{l} \right)^2$ and $\hat{S}^{x}_{l} $, see Fig.\ref{fig:jointdist}\cite{correction}. 
In the "spin diffusion regime" 
with short $l$, the joint distributions form a "ring" during the time evolution, whereas 
in the "spin decay regime" with long $l$, they  form a "disk"-like structure in the long time limit. As we will see later, a 
dimensionless parameter given by $l_{0} \sim \frac{\pi^2 l  }{4K_{s} \xi_{s}}$ determines
whether the dynamics belongs to the "spin diffusion regime"($l_{0} \leq 1$) or the "spin decay regime"($l_{0} \gg 1$).

We emphasize that in three dimensions, spin waves are dominated by $k=0$ mode 
and therefore, there is almost no decay in the magnitude of spins throughout the dynamics. 
Therefore, the existence of two qualitatively different behaviors of distribution functions of 
$\left( \hat{S}^{\perp}_{l} \right)^2$ unambiguously distinguishes the dynamics in one and three dimensions.

\subsection{Dynamics of interference between split condensates} \label{summary:interference}
\begin{figure}[t]
\begin{center}
\includegraphics[width = 7cm]{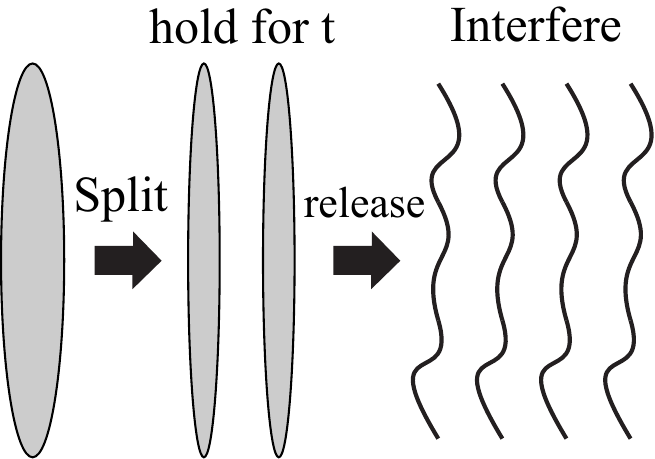}
\caption{ The interference of two quasi-condensates that are created by 
splitting one single quasi-condensate. After the splitting, the quasi-condensates
are held for time $t$ and then the transverse confinement is released. 
The two condensates interfere with each other after the release and the position of constructive 
interference is denoted by solid line in the figure. 
This interference pattern contains the information about
the local phase difference between the two quasi-condensate at the time of release. }
\label{interference}
\end{center}
\end{figure}

The dynamics of Ramsey sequence considered above can be directly mapped 
to the dynamics of interference pattern of split one dimensional 
quasi-condensate\cite{Schumm2005}. More specifically we consider the following sequence 
of operations(see Fig.\ref{interference}). First, we prepare one-component 1D quasi-condensate
in equilibrium. At time $t=0$, the quasi-condensate is quickly split along the 
axial direction, and the resulting two quasi-condensates are completely separated.
The two quasi-condensates freely evolve for a hold time of $t$, and they are 
released from transverse traps to observe the interference pattern between 
these two quasi-condensates. Such dynamics of the interference patterns 
as a function of hold time $t$ has been observed in the experiment by Hofferberth {\it et.al.}\cite{Hofferberth2007},
where the average of interference patterns is analyzed in details. Here we study 
the unique one dimensional dynamics by looking at the full distributions of interference patterns
 (for an experimental study see: Gring et al\cite{Gring2011}. ). 

The dynamics of split condensates can be mapped to the dynamics of the Ramsey interferometer studied 
in this paper. The splitting of a quasi-condensate corresponds to 
the initial $\pi/2$ pulse in the Ramsey sequence. 
If we call one of the quasi-condensates to be $L$ for left and another to be $R$ 
for right, $L$($R$)-condensate corresponds to spin-up (spin-down) component.
Thus, the density difference between the two condensates corresponds to 
 $z$ component of the spin, namely
 $\hat{S}^{z}(r,t) = \psi^{\dagger}_{L}(r,t)\psi_{L}(r,t) - \psi^{\dagger}_{R}(r,t)\psi_{R}(r,t)$
 where $\psi^{\dagger}_{i}(r,t)$ is the creation operator of particles in $i=L, R$($R$) condensate.
 Moreover, the local phase difference between the two condensates corresponds to the 
 local spin direction in $x-y$ plane. To see this, we first note that $ \psi^{\dagger}_{L}(r,t)\psi_{R}(r,t)$
 corresponds to the spin raising operator $\hat{S}^{+}(r,t)$.
 This operator is expressed in terms of the phase difference between the condensates $\hat{\phi}_{s}(r,t)$ as 
 $\hat{S}^{+}(r,t) \equiv  \psi^{\dagger}_{L}(r,t)\psi_{R}(r,t) \sim \rho e^{i \hat{\phi}_{s}(r,t)}$ where $\rho$
 is the average density of each condensate.
 Thus, for example, $x$ and $y$ spin operators are given by 
 $\hat{S}^{x}(r,t) \equiv \rho \cos( \hat{\phi}_{s}(r,t))$ and  $\hat{S}^{y}(r,t) \equiv \rho \sin( \hat{\phi}_{s}(r,t))$.
Immediately after the splitting,
the phases of the two quasi-condensates at the same coordinate along the axial direction 
are the same. 
Therefore, the splitting prepares spins in $x$ direction in the language of the Ramsey sequence, 
and thus the splitting effectively amounts to the $\pi/2$ pulse. 

The interference of two quasi-condensates measures the local phase difference 
at time $t$. If the phases of $L$ and $R$ condensates are the same, the interference pattern
has a constructive peak at the center of two condensates, which we call $x=0$. 
Thus, a shift in the interference pattern measures the local phase difference between the two condensates, which 
yields the information about $\hat{S}^{x}(r) = \rho \cos(\hat{\phi}_{s}(r))$ as well as 
$\hat{S}^{y}(r) = \rho \sin(\hat{\phi}_{s}(r))$. We note that the integrated interference contrast 
that can be extracted from experiments is given by the expression 
$\hat{C}^2 = \left| \rho \int_{l} e^{i\hat{\phi}_{s}(r)} \right|^2$\cite{Polkovnikov2006} 
and related to the transverse magnitude of spins as $\left( \hat{S}^{\perp}_{l} \right)^2 = 
\left| \hat{S}^{x}_{l} + i  \hat{S}^{y}_{l} \right|^2 = \hat{C}^2$. 

All the results obtained for Ramsey dynamics are directly applicable to the dynamics of interference 
patterns between split condensates. When the splitting process prepares two condensates with 
equal average number of particles, "spin" and "charge" degrees of freedom decouple, 
see Sec.\ref{section:interference} for details. 
In particular, depending on the integration length of the interference 
patterns along the axial direction, there exists two regimes corresponding to "phase diffusion regime" 
and "contrast decay regime," analogous to "spin diffusion regime" and "spin decay regime" described 
in the previous section, respectively. 

\subsection{Prethermalization of one-dimensional condensates} \label{summary:prethermalization}
The equilibration and relaxation dynamics of generic many-body systems are fundamental open problems. 
Among possible processes, it has been suggested that 
the time evolution of some systems prepared in non-equilibrium states  results in the reaching of quasi-steady states 
within much shorter time than equilibration time. This quasi-steady state is often not a true equilibrium state, but 
rather it is a dephased state, and true equilibration takes place at much longer time scale. 
Yet in some cases, the physical observables in the quasi-steady states 
take the value corresponding to the one in thermal equilibrium at some effective temperature $T_{\textrm{eff}}$, 
displaying disguised "thermalized" states. 
Such surprising non-equilibrium phenomena, called prethermalization, have been predicted to occur 
in quantum as well as classical many-body systems\cite{Berges2004, Barnett2010,Mathey2010} and 
observed in Gring {\em et al}\cite{Gring2011}.

In particular, integrable one dimensional systems are known {\it not to} thermalize and 
indeed, experiments in Ref.~\cite{Kinoshita2006} 
have observed an exceedingly long equilibration time. 
Yet even in this extreme case, we suggest in this section
that many-body one dimensional systems can reach disguised "thermalized" states through prethermalization 
phenomena within a short time. 

Ramsey dynamics and dynamics of interference patterns between split condensates described in the 
previous sections are particular examples of dynamics in which slow equilibration is expected because the system
essentially consists of uncoupled harmonic oscillators in the low energy description (see Sec.\ref{section:nomix}). 
In the following, we give a heuristic argument that 
the distribution of the interference contrast amplitudes of the two {\it non-equilibrium} quasi-condensates 
are given by that of two {\it equilibrium} quasi-condensates at some effective temperature $T_{\textrm{eff}}$.  
We give more details in Sec. \ref{section:prethermalization}. 

Long time after the splitting, the position of the interference peak becomes completely random, 
and yields no information. Therefore, we focus on 
the squared transverse magnitude of the spin $\left( \hat{S}^{\perp}_{l} \right)^2$, or equivalently, 
the interference contrasts $\hat{C}^2$. 
In the following, we describe the physics for the dynamics of split condensates, but the same argument 
can be applied to Ramsey dynamics. 
The interference contrast of the split condensates after a long time is determined by the 
"average" phase fluctuations present in the system. In the dephased limit, such fluctuations are 
determined by the total energy present in each mode labeled by momenta $k$. 
Now for sufficiently fast splitting, the energy $E_{\textrm{split}}$ contained in each mode 
is independent of momenta because the density difference of quasi-condensates 
along the axial direction is uncorrelated in the initial state 
beyond the spin healing length $\xi_{s}$(see discussion below Eq.~(\ref{spininistate})). 
On the other hand, the interference contrast of {\it thermal} condensates at temperature $T$ is 
determined by the thermal phase fluctuations caused by excitations whose energy is distributed 
according to equipartition theorem; each mode in the {\it thermal} condensates 
contains the equal energy of $k_{B} T$. Thus from this argument, we find that 
the interference contrast of split condensates after a long time becomes indistinguishable 
from the one resulting from thermal condensates at temperature $k_{B} T_{\textrm{eff}} = 
E_{\textrm{split}}$. We will show in Sec.\ref{section:prethermalization} that in fact, the {\it full distribution
function} of interference contrast becomes indistinguishable for these two states. 
 In the case of splitting two dimensional
condensates, equipartition of energy and existence of 
 "non-equilibrium temperature" was pointed out in Ref.\cite{Mathey2010}.

Here we propose the occurrence of such prethermalization within Tomonaga-Luttinger theory. 
We emphasize that within Tomonaga-Luttinger theory of low energy excitations, 
different modes are decoupled and therefore no true thermalization can take place. 
In realistic experimental situations, such integrability can be broken and relaxation and thermalization
process are expected to occur after a long time dynamics. 
The requirement to observe the prethermalization phenomena predicted in this theory 
in experiments depends on the time-scale of other possible thermalization processes we did not consider in our model 
such as the effective three-body collisions\cite{Mazets2008,Mazets2010}, 
relaxation of high energy quasi-particles\cite{Tan2010}, or interactions among the collective modes 
through anharmonic terms we neglected in Tomonaga-Luttinger theory\cite{Burkov2007}. 
When all these processes occur at much slower time scale than the prethermalization time-scale, which 
is roughly given by the integration length divided by the spin sound velocity $\sim l/c_{s}$ for decoupling case, 
the observation of prethermalization should be possible. In one dimension, the dynamics 
is strongly constrained due to the conservation of energy and momentum, and therefore it is likely that
the dynamics is dominated by the modes described by Tomonaga-Luttinger theory for long time 
for quasi-condensates with low initial temperatures. 

Such prethermalizations are expected to occur even in higher dimensional systems\cite{Mathey2010,Hadzibabic2009,Hadzibabic2006, Imambekov2007a}.
We note that the conditions for the experimental observations of the phenomena might be more stringent because
true thermalization processes are expected to take place much more quickly in two and three dimensional systems.

\section{Two component Bose mixtures in one dimension: Hamiltonian} 
\label{section:hamiltonian}
In this paper, we study the dynamics of 
two-component Bose mixtures in one dimension through Tomonaga-Luttinger 
formalism\cite{Giamarchi2004,Cazalilla2011}. As we have stated before, we assume the rotating frame, 
in which spin-up and spin-down particles have the same chemical potential
in the absence of interactions.
The  Hamiltonian of two component Bose mixtures in one dimension is given by
\begin{eqnarray}
{\cal H} =  \, \int_{-L/2}^{L/2} dr \left[   \, \sum_i 
 \frac{1}{2m_{i}} \nabla \psi^{\dagger}_i(r) \nabla
\psi_i(r) \, \right. \nonumber \\ 
\left. + \, \sum_{ij} g_{ij}  \, 
\psi^{\dagger}_{i}(r) \psi^{\dagger}_{j}(r)
\psi_{j}(r)\psi_{i}(r) \right]
\label{microscopic}
\end{eqnarray}
Here $\psi_{i}$ with $i=\uparrow, \downarrow$ describe two atomic species with
masses $m_{i}$ and $g_{ij}$ are the interaction strengths given by
$g_{ij} = \nu_{\perp} a_{ij}$\cite{Olshanii1998} where $\nu_{\perp}$ is the frequency of transverse confinement potential
and $a_{ij}$ are the scattering lengths between spin $i$ and $j$. 
System size is taken to be $L$, and we take the periodic boundary condition throughout the paper. 
In addition, we use the units in which $\hbar =1$. 

In the low energy description, the Hamiltonians for weakly interacting bosons after
the initial $\pi/2$ rotation can be written in quadratic form, and given by
\begin{eqnarray}
H & = & H_{\uparrow} + H_{\downarrow} + H_{int}, \label{hamiltonian} \\
H_{\uparrow} &=& \int^{L/2}_{-L/2} dr 
\left[ \frac{\rho}{2m_{\uparrow}} (\nabla \hat{\phi}_{\uparrow}(r))^2 
+ g_{\uparrow\uparrow} (\hat{n}_{\uparrow}(r))^2 \right], \nonumber \\
H_{\downarrow} &=& \int^{L/2}_{-L/2} dr 
\left[ \frac{\rho}{2m_{\downarrow}} (\nabla \hat{\phi}_{\downarrow}(r))^2 
+ g_{\downarrow \downarrow}(\hat{n}_{\downarrow}(r))^2 \right], \nonumber \\
H_{int} & = & 2 \int^{L/2}_{-L/2} dr 
 (r)\left[g_{\uparrow \downarrow} \hat{n}_{\uparrow}\hat{n}_{\downarrow}(r) +
 g^{\phi}_{\uparrow \downarrow} \nabla \hat{\phi}_{\uparrow}
 \nabla \hat{\phi}_{\downarrow}(r)\right],
  \nonumber
\end{eqnarray}
where $\rho$ is the average density of each species 
and $\hat{n}_{\sigma}$ are variables representing the phase and density fluctuation
for the particle with spin $\sigma$. These variables obey a canonical commutation relation 
$[\hat{n}_{\sigma}(r),\hat{\phi}_{\sigma}(r')]= -i \delta(r-r')$. 
 In the Hamiltonian above, we included the kinetic interaction term 
$g^{\phi}_{\uparrow \downarrow}$, which is zero for weakly interacting bosons,
but allowed by inversion symmetry and non-zero for generic Tomonaga-Luttinger Hamiltonians. 

We note that in the weakly interacting case, one can obtain the parameters of 
the Hamiltonian in Eq. (\ref{hamiltonian}) such as $g_{ij}$ and $m_{i}$ through 
hydrodynamic linearization of the microscopic Hamiltonian. In this case, we assume the small 
fluctuations of the densities $\hat{n}_{\sigma}$ and phases $\nabla \hat{\phi}_{\sigma}$ 
and expand the Hamiltonian in Eq.~(\ref{microscopic}) to the second order in these variables 
through the expression $\psi^{\dagger}_{\sigma} \sim \sqrt{\rho + \hat{n}_{\sigma}} e^{i\hat{\phi}_{\sigma}}$,
resulting in the form of the Hamiltonian in Eq. (\ref{hamiltonian}). Due to this assumptions of 
small spatial variations of the phase, $\hat{n}_{\sigma}$ and $\hat{\phi}_{\sigma}$
represent the ''coarse-grained'' variables where collective modes have linear dispersions. 
The Hamiltonian of Tomonaga-Luttinger theory in Eq.~(\ref{hamiltonian}) can also describe
effective low-energy physics of strongly interacting systems, but in this case
there is no simple relation between microscopic parameters
and the parameters of the Hamiltonian in Eq.~(\ref{hamiltonian}). 


In order to describe the spin dynamics, we define 
spin and charge operators as the difference and the sum 
of spin up and down operators, {\it i.e.}
$\hat{\phi}_{s} = \hat{\phi}_{\uparrow} - \hat{\phi}_{\downarrow}$,
$\hat{\phi}_{c} = \hat{\phi}_{\uparrow} + \hat{\phi}_{\downarrow}$,
$\hat{n}_{s} = \frac{1}{2}(\hat{n}_{\uparrow} - \hat{n}_{\downarrow})$,
$\hat{n}_{c} = \frac{1}{2}(\hat{n}_{\uparrow} + \hat{n}_{\downarrow})$.
In this representation,  
the Hamiltonian in Eq. (\ref{hamiltonian}) becomes
\begin{eqnarray}
H & = & H_{s} + H_{c} + H_{mix} \nonumber \\
H_{s} &=& \int^{L/2}_{-L/2} dr 
\left[ \frac{\rho}{2m_{s}} (\nabla \hat{\phi}_{s}(r))^2 + g_{s} (\hat{n}_{s}(r))^2 \right]
\label{spinhamiltonian0}\\
H_{c} &=& \int^{L/2}_{-L/2} dr 
\left[ \frac{\rho}{2m_{c}} (\nabla \hat{\phi}_{c}(r))^2 + g_{c} (\hat{n}_{c}(r))^2 \right]
\label{chargehamiltonian0}\\
H_{mix} & = & 2  \int^{L/2}_{-L/2} dr \left[ g_{mix}   \hat{n}_{s}(r)\hat{n}_{c}(r) 
+ g^{\phi}_{mix}  \nabla \hat{\phi}_{s}(r) \nabla \hat{\phi}_{c}(r) \right] \nonumber \\
&& \label{mixinghamiltonian}
\end{eqnarray}
where interaction strengths are given by
$g_{c} = g_{\uparrow \uparrow} + g_{\downarrow \downarrow} 
+ 2 g_{\uparrow \downarrow}$,
$g_{s} = g_{\uparrow \uparrow} + g_{\downarrow \downarrow} 
- 2 g_{\uparrow \downarrow}$,
$g_{mix} = g_{\uparrow \uparrow} - g_{\downarrow \downarrow}$, 
$g_{mix}^{\phi} =\rho /(8m_{\uparrow}) - \rho/(8m_{\downarrow} )$.
The masses are given by the relations 
$\rho/(2m_{c})=\rho /(8m_{\uparrow}) + \rho/(8m_{\downarrow} )
+ g^{\phi}_{\uparrow \downarrow}/2$ and 
$\rho/(2m_{s})=\rho /(8m_{\uparrow}) + \rho/(8m_{\downarrow} )
- g^{\phi}_{\uparrow \downarrow}/2$.

 The spin variables $\hat{\phi}_{s}$ and $\hat{n}_{s}$ are 
"coarse-grained" in the sense that they represent the operators in the 
long wavelength beyond the spin healing length $\xi_{s}$. $\xi_{s}$ 
is determined from microscopic physics and gives the length below which 
the kinetic energy of spins wins over the interaction energy, see Eq.~(\ref{spinhamiltonian0}) above. 
For weakly interacting bosons with 
$m_{\uparrow} = m_{\downarrow}= m$, it is given by 
$\xi_{s}= \pi /\sqrt{m \rho g_{s}}$. 
In the following, we assume that the number 
of particles within the spin healing length is large, {\it i.e.} $ \xi_{s} \rho \gg 1$.
This condition is always satisfied for weakly interacting bosons. 


In the next section, we consider the case $g_{mix} = 0$ and $g^{\phi}_{mix} =0$,
in which spin and charge degree of freedom decouple. Then the dynamics 
of spins is completely described by the spin Hamiltonian in Eq. (\ref{spinhamiltonian0}).
The general case in which $g_{mix} \neq 0$ and $g^{\phi}_{mix} \neq 0$
will be treated in Sec~\ref{section:mix}.

\section{Dynamics of Full Distribution Function 
for decouped spin and charge degrees of freedom} \label{section:nomix}
\subsection{Hamiltonian and initial state}
The experiment of Widera {\it et al.}\cite{Widera2008} used  $F=1, m_{F}=+1$ and 
 $F=2, m_{F}=-1$ states of $^{87}$Rb for spin-up and spin-down particles, respectively. 
 These hyperfine states have the scattering lengths $a_{\sigma \sigma'}$ such that 
 $a_{\uparrow \uparrow} \approx a_{\downarrow \downarrow}$.
 Consequently, the mixing Hamiltonian in Eq.(\ref{mixinghamiltonian}) approximately vanishes for weak interactions. 
Motivated by this experiment, here 
we consider the decoupling of spin and charge degrees of freedom\cite{Kitagawa2010}.
Spin dynamics in this case is completely determined by the spin 
Hamiltonian 
\begin{eqnarray}
H_{s} & =& \frac{c_{s}}{2} \int \, \left[ \, \frac{K_{s}}{\pi} (\nabla \hat{\phi}_{s}(r) )^2 +
\frac{\pi}{K_{s}} \hat{n}_{s}^2(r) \, \right] \, dr
 \label{spinhamiltonian}
\end{eqnarray}
where $K_{s}$ is the spin Luttinger parameter
representing the strength of interactions, and $c_{s}$ is spin sound velocity.
$K_{s}$ and $c_{s}$ are directly related to the spin healing length $\xi_{s}$
in the weak interaction limit, given by $2K_{s} = \rho \xi_{s}$ and $c_{s} = \frac{\pi}{2m_{s} \xi_{s}}$.
$\hat{n}_{s}(r,t)$ is the local spin imbalance $\hat{n}_{s} =
\psi^\dagger_\alpha (\frac{1}{2} \sigma^z_{\alpha\beta}) \psi_\beta $ and
$\hat{\phi}_{s}(r,t)$ is related to the direction of the transverse spin component
$\rho e^{i\hat{\phi}_{s}} = \psi^\dagger_\alpha \sigma^+_{\alpha\beta} \psi_\beta$.
Here, $\psi^\dagger_\alpha$ is the creation operator of spin $\alpha =\uparrow, \downarrow$. 
These variables $\hat{n}_{s}$ and $\hat{\phi}_{s}$ obey a canonical commutation relation
$[\hat{n}_{s}(r),\hat{\phi}_{s}(r')]= -i \delta(r-r')$.

Other spin variables can be similarly defined in terms of coarse grained spin variables 
$\hat{n}_{s}$ and $\hat{\phi}_{s}$. In the following, we consider the general transverse spins 
pointing in the direction $(x,y,z) = (\cos\theta, \sin\theta, 0)$ integrated over $l$
given by
\begin{eqnarray}
\hat{S}^{\theta}_{l} &=&
 \int_{-l/2}^{l/2} dr \, \psi^\dagger_\alpha (r,t)
\left( \cos\theta \frac{\sigma^x_{\alpha\beta}}{2} + \sin\theta \frac{\sigma^y_{\alpha\beta}}{2} 
\right) \psi_\beta (r,t)  \nonumber \\
& = & \int^{l/2}_{-l/2} dr \frac{\rho}{2} 
\left( e^{i(\hat{\phi}_{s}(r) - \theta)} + e^{-i(\hat{\phi}_{s}(r) - \theta)}\right)
\label{spinvariables} 
\end{eqnarray}
where $\sigma^{a}$ with $a=x,y$ are Pauli matrices. 
Here $\hat{S}^{\theta}_{l}$ with $\theta =0$ corresponds to spin $x$ operator
and $\theta = \pi/2$ corresponds to spin $y$ operator. 
In order to explore the one dimensional dynamics resulting from Hamiltonian
in Eq.(\ref{spinhamiltonian}), we analytically compute
the $m$th moment of the spin operator $\hat{S}^{\theta}_{l}$, 
$\braket{\left( \hat{S}^{\theta}_{l} \right)^m}$, after time $t$
of the $\pi/2$ pulse of the Ramsey sequence. Then, the full distribution functions 
of $\hat{S}^{x}_{l}$ and $\hat{S}^{y}_{l}$, as well as the joint distribution of these will be 
obtained from $\braket{\left( \hat{S}^{\theta}_{l} \right)^m}$. 

In order to study the dynamics of Ramsey interferometer in terms of 
low energy variables $\hat{n}_{s}$ and $\hat{\phi}_{s}$, we need to write down 
an appropriate state after the $\pi/2$ pulse in terms of $\hat{n}_{s}$ and $\hat{\phi}_{s}$. 
If pulse is sufficiently 
strong, each spin is independently rotated into $x$ direction after the $\pi/2$ pulse.
Naively, this prepares the initial state in the eigenstate of 
$\hat{S}^{x}(r)= \rho \cos\hat{\phi}_{s}(r)$ with eigenvalue $\hat{\phi}_{s}(r)=0$. 
However, due to the commutation relation between $\hat{n}_{s}$ and $\hat{\phi}_{s}$, 
such an initial state has an infinite fluctuation in $\hat{n}_{s}$ and therefore, the state 
has an infinite energy according to Eq.(\ref{spinhamiltonian}). 
This unphysical consequence comes about because the low energy theory
in Eq.(\ref{spinhamiltonian}) should not be applied to the physics of short time scale
given by $1/E_{c}$ where $E_{c}$ is the high energy cutoff of Tomonaga-Luttinger theory. 
During this short time dynamics, the initial state establishes the correlation at the
length scale of spin healing length $\xi_{s}$. The state after this short time dynamics
can now be described in terms of the coarse-grained variables $\hat{n}_{s}(r)$ and $\hat{\phi}_{s}(r)$. 
The variables $\hat{n}_{s}(r)$ and $\hat{\phi}_{s}(r)$ are defined on the length scale larger 
than the spin healing length $\xi_{s}$. Since the $z$ component of spins are 
still uncorrelated beyond $\xi_{s}$ after the initial short time dynamics, the appropriate initial 
condition of the state is written as 
\begin{equation}
\langle  S^{z}(r) S^{z}(r') \rangle =\langle  \hat{n}_{s}(r) \hat{n}_{s}(r') \rangle = \frac{\rho \eta}{2} \delta(r-r') \label{spininistate}
\end{equation}
where the delta function $\delta(r-r')$ should be understood as a smeared delta function 
over the scale of $\xi_{s}$. Because the state after the short time dynamics is still 
close to the eigenstate of the $\hat{S}^{x}(r)$ operator, spins are equal superpositions of spin-up and spin-down. 
Then the distribution of $\hat{S}^{z}_{l} = \int^{l}_{0} \hat{S}^{z}(r) dr $ is
determined through random picking of the values $\pm 1/2$ for $2\rho l$ particles. 
Due to the central limit theorem, the distribution of $\hat{S}^{z}_{l} = \int^{l}_{0} \hat{S}^{z}(r) dr $ is Gaussian, {\it i.e.}
$\braket{\left(\hat{S}_{l}^{z}\right)^{2n}} = \frac{(2n)!}{2^n n!} \left( \rho l \eta \right/2)^n $. In particular, 
$\braket{\left(\hat{S}^{z}_{l}\right)^2} = \rho l \eta /2$, which determines the magnitude of the fluctuation 
for $\hat{S}^{z}(r)$ in Eq.(\ref{spininistate}). 
In Eq.(\ref{spininistate}), we also introduced the phenomenological parameter $\eta$ which accounts for 
the decrease and increase of fluctuations coming from, for example, imperfections
of $\pi/2$ pulse. The ideal, fast application of $\pi/2$ pulse  corresponds to $\eta =1$. 
 In the experimental realization of Ref.~\cite{Widera2008}, 
$\eta$ was determined to be between $0.8$ and $1.3$ through the fitting 
of the experiment with Tomonaga-Luttinger theory 
for the time evolution of the average $x$ component of the spin, $\braket{\hat{S}^{x}_{l}}$.
Through engineering of the initial state such 
as the application of a weak $\pi/2$ pulse, $\eta$ can also be made intentionally smaller than $1$. 

A convenient basis to describe the initial state of the dynamics above is the basis that diagonalizes
the spin Hamiltonian of Eq.(\ref{spinhamiltonian}). The phase and density of the spins
$\hat{\phi}_{s}(r)$ and $\hat{n}_{s}(r)$ can be written in terms of 
the creation $b_{s,k}^{\dagger}$ and annihilation $b_{s,k}$ operators
of elementary excitations for the spin Hamiltonian in Eq.(\ref{spinhamiltonian}) as 
\begin{eqnarray}
\hat{\phi}_{s}(r) 
& = & \frac{1}{\sqrt{L}} \sum_{k} \hat{\phi}_{s,k} e^{ikr} \nonumber \\
& = & \frac{1}{\sqrt{L}}  \left(\sum_{k \neq 0}-i \sqrt{ \frac{\pi}{2|k|K_{s}}}(b_{s,k}^{\dagger} - b_{s,-k}) e^{ikr} + \hat{\phi}_{s,0} \right)  \nonumber \\
\hat{n}_{s}(r)  
& = & \frac{1}{\sqrt{L}} \sum_{k} \hat{n}_{s,k} e^{ikr} \nonumber \\
& = &  \frac{1}{\sqrt{L}}  \left(\sum_{k \neq 0}\sqrt{ \frac{|k|K_{s}}{2\pi }}(b_{s,k}^{\dagger} + b_{s,-k}) e^{ikr} + \hat{n}_{s,0} \right)\nonumber \\
{\cal H}_{s} &=& \sum_{k \neq 0} c_{s} |k| b_{s,k}^\dagger b_{s,k} 
+ \frac{\pi c_{s}}{2 K_{s}} \, \hat{n}_{s,0}^2 \label{spindiagonalhamiltonian}
\end{eqnarray} 
\newline
where we defined $\hat{\phi}_{s,k}$ and $\hat{n}_{s,k}$ to be the Fourier transform of 
operators $\hat{\phi}_{s}(r)$ and $\hat{n}_{s}(r)$. 
$b^{\dagger}_{s,k}$ creates a collective mode with momentum $k$ and follows a
canonical commutation relation $[b_{s,k}, b^{\dagger}_{s,k}]=1$. 
Note that $k=0$ mode has no kinetic energy, and it naturally has different 
evolution from $k \neq 0$ modes. 

The Gaussian state determined by Eq.(\ref{spininistate}) takes the form of
a squeezed state of operators $b_{s,k}$, and it is given by
\begin{eqnarray}
| {\psi_{0}}  \rangle &=& \frac{1}{\mathcal{N}} \exp{\left(\sum_{k\neq 0} W_{k} b_{s,k}^{\dagger} b^{\dagger}_{s,-k} \right)} | {0} \rangle
| {\psi_{s,k=0}} \rangle
\nonumber \\
\langle n_{s,0} | {\psi_{s,k=0}} \rangle & = & \exp\left(-\frac{1}{2\rho \eta} n_{s,0}^2\right) \label{spininitial_state} 
\end{eqnarray} 
where $2W_{k} =  \frac{1-\alpha_{k}}{1+\alpha_{k}}$, 
$\alpha_{k} = \frac{|k|K_{s}}{\pi \rho \eta} $.
Here the state $\ket{n_{s,0}}$ is the normalized eigenstate of the operator $\hat{n}_{s,0}$
with eigenvalue $n_{s,0}$. 
The summation of $k$ in the exponent has a ultraviolet cutoff around $ k_{c} = 2\pi/\xi_{s}$. 
$\mathcal{N}$ is the overall normalization of the state.
It is easy to check that $\bra{\psi_{0}} \hat{n}_{s,k} \hat{n}_{s,k'} \ket{\psi_{0}}
= \frac{\rho \eta}{2} \delta_{k,-k'}$, which corresponds to Eq. (\ref{spininistate}).

\subsection{Moments and full distribution functions of spins}\label{section:nomixing}
After free evolution of the initial state $\ket{\psi_{0}}$ for time $t$, the state becomes
$\ket{\psi(t)} = e^{-iH_{s}t} \ket{\psi_{0}}$. We characterize the state at time $t$ by 
the $m$th moments of spin operators,$\braket{(\hat{S}_{l}^{\theta})^m}$.  
As we will see below, the full distribution function can be constructed 
from the expression of $\braket{(\hat{S}_{l}^{\theta})^m}$ \cite{Imambekov2008a}.

We consider the evaluation of moments $\braket{(\hat{S}_{l}^{\theta})^m}$ at 
time $t$, $\ket{\psi(t)}$. Each momentum $k$ component of 
the initial state $| {\psi_{0}}  \rangle$ independently evolves in time. 
Since $k=0$ mode has a distinct evolution from other $k \neq 0$ 
modes, we separately consider $k=0$ and $k \neq 0$ modes.

\subsubsection{$k=0$ mode}
The Hamiltonian of $k=0$ mode is given by 
$H_{s,k=0} = \frac{\pi c_{s}}{2 K_{s}} \, \hat{n}_{s,0}^2$ 
in Eq. (\ref{spindiagonalhamiltonian}). 
Therefore, 
in the basis of $n_{s,0}$, $k=0$ part of the state $\ket{\psi(t)}$ is given by
\begin{eqnarray}
&\bra{n_{s,0}} e^{-iH_{s,k=0}t} \ket{\psi_{k=0}} = \nonumber \\
&\frac{1}{\mathcal{N}_{k=0}} \exp{\left\{ 
\left(-\frac{1}{(2\rho \eta )} -i \frac{\pi c_{s}t}{2 K_{s}}\right) n_{s,0}^2 \right\}},
\end{eqnarray}
where $\mathcal{N}_{k=0}$ is the normalization of the state.
The initial Gaussian state of $\hat{n}_{s,0}$ stays Gaussian at all times, 
and any analytic operator of $\phi_{s,0}$ and $n_{s,0}$ 
can be exactly evaluated through Wick's theorem. 
For example, $k=0$ part contributes to the decay of the average of the $x$ component
of spin $\braket{\hat{S}^{x}_{l}}_{k=0} = l\rho$Re$\left(\braket{e^{i\phi_{s,0}/\sqrt{L}}}\right)$ as 
\begin{eqnarray}
\braket{\hat{S}^{x}_{l}}_{k=0} &=& 
l\rho e^{-\frac{1}{2L} \braket{\phi_{s,0}^2}_{t}} \nonumber \\
\braket{\phi_{s,0}^2}_{t} & = & \frac{1}{2\rho \eta} + \left(\frac{c_{s}\pi t}{K_{s}}\right)^2 
 \frac{\eta \rho}{2}
\label{zeromomentum}
\end{eqnarray}
This diffusion of the spin from $k=0$ contribution
is generally present in any dimensional systems, and not particular to 
one dimension. Physical origin of this diffusion is the interaction dependent on the total spin
, $\hat{S}_{z}^2$. The eigenstate of $\hat{S}_{x}$ with eigenvalue $\rho l$ is the superposition of 
different eigenstates of $\hat{S}_{z}$ with eigenvalues $m_{z}$, 
and they accumulate different phases $e^{-itm_{z}^2}$ in time. 
This leads to the decay of $\braket{\hat{S}_{x}}$. In the thermodynamic limit $L \rightarrow \infty$, 
the uncertainty of $\hat{S}_{z}$ becomes diminishingly small, and therefore, the decay of $\braket{\hat{S}_{x}}$
coming from $k=0$ goes to zero. 
More interesting physics peculiar to one dimensional systems 
comes from $k \neq 0$ modes. In the case of three dimensional systems,
macroscopic occupancy of a single particle state is absent in one dimension,
so $k \neq 0$ momentum excitations have much more significant effect
 in one dimensional dynamics. 

\subsubsection{$k \neq 0$ contribution}
The exact evaluation of spin moments 
$\braket{(\hat{S}_{l}^{\theta})^m}$ for $k \neq 0$ is possible 
through the following trick.
Consider the annihilation operator $\gamma_{s,k}(t)$ 
for the state $\ket{\psi(t)}$ 
such that $\gamma_{s,k}(t) \ket{\psi(t)} = 0$. 
If we write the operators $\hat{\phi}_{s}(r)$ in terms of 
$\gamma_{s,k}(t)$ and $\gamma_{s,k}^{\dagger}(t)$, 
then $k \neq 0$ part of the $m$th moment 
schematically takes the form
$\braket{(\hat{S}_{l}^{\theta})^m} \sim \braket{\exp( 
i\sum_{k\neq 0 } C_{s,k} \gamma_{s,k} + C_{s,k}^{*} \gamma_{s,k}^{\dagger}) }$ 
(Here and in the following, we drop the time dependence of $\gamma_{s,k}(t)$ from the notation).
Using the property 
$e^{\gamma_{s,k}}\ket{\psi(t)} = (1+\gamma_{s,k} + \cdots)\ket{\psi(t)} = \ket{\psi(t)}$ and the identity $e^{A+B} = e^{A}e^{B}e^{-\frac{1}{2}[A,B]}$
where $[A,B]$ is a c-number, we can evaluate $m$th moments as  
$\braket{(S_{l}^{\theta})^m} \sim 
\braket{e^{iC_{s,k}^{*}\gamma_{s,k}^{\dagger}}  e^{-\frac{1}{2} \sum_{k\neq0} |C_{s,k}|^2} e^{iC_{s,k} \gamma_{s,k}}} = \exp(-\frac{1}{2} \sum_{k\neq0} |C_{s,k}|^2)$.

It is straightforward to check that $\gamma_{s,k}$ operator is given by 
the linear combination of $b_{s,k}$ and $b_{s,-k}^\dagger$ as follows, 
\begin{equation}
 \left( \begin{array}{c} \gamma_{s,-k}^{\dagger}(t) \\ \gamma_{s,k}(t) \end{array}\right) 
  =  
 \left( \begin{array}{cc} 
 \frac{e^{-ic_{s}|k|t}}{\sqrt{1-4|W_{k}|^2}} 
 & \frac{-2W_{k}e^{ic_{s}|k|t}}{\sqrt{1-4|W_{k}|^2}}   \\ 
  \frac{-2W_{k}e^{-ic_{s}|k|t}}{\sqrt{1-4|W_{k}|^2}} 
  & \frac{e^{ic_{s}|k|t}}{\sqrt{1-4|W_{k}|^2}}
  \end{array}\right) 
 \left( \begin{array}{c} b_{s,-k}^{\dagger} \\ b_{s,k} \end{array}\right). \label{gamma}
\end{equation}
$\gamma_{s,k}$ and $\gamma_{s,k}^\dagger$ obey a canonical commutation relation 
$[\gamma_{s,k}, \gamma_{s,k}^{\dagger}] = 1$. 
In terms of these $\gamma_{s,k}$, the expression of $\hat{\phi}_{s,k}(t)$ 
becomes 
 \begin{eqnarray}
 \frac{1}{\sqrt{L}} \hat{\phi}_{s,k} &=& C_{s,k}\gamma_{s,k}^{\dagger} + C_{s,k}^{*}\gamma_{s,-k} \nonumber \\
 C_{s,k} &=& -i\sqrt{ \frac{\pi}{2|k|K_{s}L}}
  \frac{e^{ic_{s}|k|t}-2W_{k}e^{-ic_{s}|k|t}}{\sqrt{1-4|W_{k}|^2}}. \label{ck}
 \end{eqnarray} 
 $C_{s,k}(t)$ measures the fluctuation, or variance, of phase in the $k$th mode at time $t$,
 given by $\braket{|\hat{\phi}_{s,k}(t)|^2}$ = $\braket{\hat{\phi}_{s,k}(t) \hat{\phi}_{s,-k}(t)}$. 
Indeed, since $\gamma_{s,k}$ is the annihilation operator of our state at time $t$, we immediately 
conclude that $\braket{|\hat{\phi}_{s,k}(t)|^2}/L = |C_{s,k}(t)|^2$. 

Using the technique described above, $m$th moment of $\hat{S}^{\theta}_{l}$
becomes (we include both $k=0$ and $k\neq 0$ contributions in the expression below)
\begin{flalign}
&\bra{\psi(t)} \left( \int^{l/2}_{-l/2} S^{\theta}(r) dr \right)^m \ket{\psi(t)} \nonumber \\
= & \braket{\prod_{i}^{m} \left( \int^{l/2}_{-l/2} \frac{\rho}{2} dr_{i} 
\sum_{s_{i} = \pm 1}e^{is_{i}(\hat{\phi}_{s}(r_{i})-\theta)} \right) } \nonumber \\
=&  \sum_{\{ s_{i} = \pm 1 \}} \prod_{i=1}^{m} \int^{l/2}_{-l/2} \frac{\rho dr_{i}}{2} 
\braket{e^{\left(i(s_{1}\hat{\phi}_{s}(r_{1}) +  \ldots +s_{m}\hat{\phi}_{s}( r_{m}) \right) }} \nonumber \\
&\times e^{-i \left(\sum_{i} s_{i}\right) \theta} \nonumber \\
%
= & \sum_{\{ s_{i} = \pm 1 \}} \prod_{i=1}^{m} \int^{l/2}_{-l/2}  \frac{\rho dr_{i}}{2}
\exp{ \left( -\frac{1}{2}\sum_{k} \xi_{s,k}^{\{s_{i},r_{i}\}} (\xi_{s,k}^{\{s_{i},r_{i}\}})^{*} \right)} \nonumber \\ 
&\times e^{-i\left(\sum_{i} s_{i}\right) \theta},
\label{complicatedmoment}
\end{flalign}
where $\xi_{s,k}^{\{s_{i},r_{i}\}} = 
\sqrt{\frac{\braket{|\hat{\phi}_{s,k}(t)|^2}}{L}}(s_{1} e^{ikr_{1}} + \ldots + s_{m} e^{ikr_{m}}) $. 
$s_{i}$ takes either the value $1$ or $-1$, 
and $\sum_{\{s_{i}\}}$ sums over all possible 
set of values. Note that $L$ is the total system size and $l$ is the integration range. 

\subsubsection{Full Distribution Function}
Calculation of the full distribution functions from moments 
in Eq. (\ref{complicatedmoment}) is studied 
by the techniques introduced in Ref.~\cite{Imambekov2008a} through mapping to the statistics of random surfaces. 
In this subsection, we provide the details of the calculation. 

Eq. (\ref{complicatedmoment}) is simplified if the integrations for 
each $r_{i}$ can be independently carried out. This is not possible in 
Eq. (\ref{complicatedmoment}) because $e^{ikr_{i}}$ and $e^{ikr_{j}}$ 
for $i \neq j$ are coupled in 
$ \left| \xi_{s,k}^{\{s_{i},r_{i}\}} \right|^2 
= \left(\textrm{Re}\xi_{s,k}^{\{s_{i},r_{i}\}}\right)^2+ 
\left(\textrm{Im}\xi_{s,k}^{\{s_{i},r_{i}\}}\right)^2$.
To unentangle this, we introduce Hubbard-Stratonovich transformation, 
$e^{-\frac{x^2}{2}} = \frac{1}{\sqrt{2\pi}} \int^{\infty}_{-\infty} 
e^{-\frac{\lambda^2}{2}} e^{ix\lambda}$, for example,
\begin{displaymath}
 e^{-\frac{1}{2} \left(\textrm{Re}(\xi_{s,k}^{\{s_{i},x_{i}\}} )\right)^2}
= \int^{\infty}_{-\infty} \frac{d\lambda_{1sk}}{\sqrt{2\pi}}  e^{-\lambda_{1sk}^2/2} e^{i\lambda_{1sk} 
\textrm{Re} \left(\xi_{s,k}^{\{s_{i},x_{i}\}}\right) }.
\end{displaymath}
We apply a similar transformation for Im$\xi_{s,k}$.
This removes the cross term between $e^{ikr_{i}}$ and $e^{ikr_{j}}$ for $i \neq j$
and allows us to independently integrate over $r_{i}$'s. 
 Associated with each transformation, 
 we introduce auxiliary variables $\lambda_{1sk}$ for Re$(\xi_{s,k})$, $\lambda_{2sk}$ for 
 Im$(\xi_{s,k})$.
 Then, $m$th moment becomes
 \begin{widetext}
 \begin{flalign}
& \bra{\psi(t)} \left( \int^{l/2}_{-l/2} 
S^{\theta}(r) dr \right)^m \ket{\psi(t)} = \nonumber \\ 
& \sum_{\{ s_{i} \}}  \prod_{k }  \int^{\infty}_{-\infty}e^{-(\lambda_{1sk}^2 + \lambda_{2sk}^2)/2} 
 \frac{d\lambda_{1sk}}{\sqrt{2\pi}}  \frac{d\lambda_{2sk}}{\sqrt{2\pi}}  \left[ \prod_{i=1}^{m} \int^{l/2}_{-l/2} \frac{\rho dr_{i}}{2} \exp \left(i s_{i} \sum_{k} 
\left\{ \lambda_{1sk} \textrm{Re}(\xi_{s,k}^{r_{i}}) + \lambda_{2sk} \textrm{Im}(\xi_{s,k}^{r_{i}})
 -\theta\right\} \right) \right],  \nonumber 
\end{flalign}
\end{widetext}
where we introduced $\xi_{s,k}^{\{r_{i}\}} 
= \sqrt{\frac{\braket{|\hat{\phi}_{s,k}(t)|^2}}{L}} e^{ikr_{i}}$. 
Summation over $\{s_{i} = \pm 1\}$ can now be carried out. 
Furthermore, we introduce a new variables $\lambda_{rsk}$ and $\lambda_{\theta sk}$,
and replace $\lambda_{1sk}$ and $\lambda_{2sk}$ through the relation 
$ \lambda_{rsk} = \sqrt{\lambda_{1sk}^2 +  \lambda_{2sk}^2}$ and 
$\cos(\lambda_{\theta sk}) = \lambda_{2sk}/\sqrt{\lambda_{1sk}^2 +  \lambda_{2sk}^2}$. 
These operations result in the simplified expression,
\begin{flalign}
&  \bra{\psi(t)} \left( \int^{l/2}_{-l/2}
 S^{\theta}(r) dr \right)^m \ket{\psi(t)} =  \prod_{k,a={r,\theta}} \frac{1}{2\pi}  \int  d\lambda_{ask} \nonumber \\
&
\times  \lambda_{rsk}e^{-\frac{\lambda_{rsk}^2}{2}} 
\left( \rho \int^{l/2}_{-l/2} dr \cos\left[\chi(r, \{\lambda_{jsk} \}) -\theta \right] \right)^m,
\label{nthmoment} 
\end{flalign}
where
\begin{flalign}
\chi(r, \{\lambda_{jsk} \})  =& \sum_{k} \sqrt{\frac{\braket{|\hat{\phi}_{s,k}|^2}} {L}}
 \lambda_{rsk} \sin(kr +\lambda_{\theta sk} ),\label{chi} 
 \end{flalign}
 \begin{flalign}
  \braket{|\hat{\phi}_{s,k}|^2}  = & 
   \frac{\pi}{2|k|K_{s}}\frac{\sin^2(c_{s}|k|t) 
+\alpha_{k}^2\cos^2(c_{s}|k|t)}{\alpha_{k}} (k \neq 0),
\nonumber \\
\braket{\phi_{s,0}^2}_{t}  = &\frac{1}{2\rho \eta} + \left(\frac{c_{s}\pi t}{K_{s}}\right)^2 
 \frac{\eta \rho}{2} \quad (k = 0),
\label{kfluctuation}
\end{flalign} 
with $\alpha_{k} = \frac{|k|K_{s}}{\pi \rho \eta}$. The integration over $\lambda_{rsk}$
and $\lambda_{\theta sk}$ in Eq.(\ref{nthmoment}) extends from $0$ to $\infty$
and from $-\pi$ to $\pi$, respectively. 

Comparing the expression in Eq.(\ref{nthmoment}) and 
the implicit definition of a distribution function in Eq.(\ref{moments}),
it is easy to identify the distribution function as 
\begin{flalign}
P^{\theta}_{l}(\alpha) = &\prod_{k} \int^{\pi}_{-\pi} \frac{d\lambda_{\theta sk}}{2\pi} \int^{\infty}_{0} \lambda_{rsk}e^{-\lambda_{rsk}^2/2} d\lambda_{rsk} \nonumber \\
\times& \delta\left(\alpha - \rho \int^{l/2}_{-l/2} dr \cos\left[\chi( r,\{\lambda_{jsk} \})-\theta \right] \right). \label{disttheta}
\end{flalign}
This function can be numerically evaluated through Monte Carlo method with weight $\lambda_{rsk}e^{-\lambda_{rsk}^2/2}$
for $\lambda_{rsk}$ and equal unity weight for $\lambda_{\theta sk}$. 

While we have assumed that the chemical potentials of spin-up and spin-down atoms 
are the same in the absence of interactions by going to the rotating frame, it is easy to 
obtain  the expression for distribution functions in the lab frame. 
The energy difference $E$ between spin-up and spin-down atoms results in the rotation of the spin
in the $x-y$ plane at a constant angular velocity $E$. Therefore, 
the distribution in the lab frame is obtained by replacing $\theta \rightarrow \theta+Et$ 
in Eq. (\ref{disttheta}). 

In this section, we have focused on the distribution function of spins in $x-y$ plane, 
but it is also possible to obtain the distribution function of $z$ component of the spin,
and we present the result in the Appendix \ref{zdist}. 

\subsubsection{Joint Distribution Function} \label{section:jointdist}
From the expression for the spin operators in Eq. (\ref{spinvariables}), we observe that 
the spin operators for the $x$ and $y$ directions commute in the low energy description. 
This is because spin operators in Tomonaga-Luttinger theory are coarse-grained over $\sim \rho \xi_{s}$ particles, 
and since for weak interactions $\rho \xi_{s} \gg 1$, the uncertainty of measurements coming from non-commutativity of 
$\hat{S}^{x}_{l}$ and $\hat{S}^{y}_{l}$ becomes suppressed. 
The possibility of simultaneous measurements of spin $x$ and $y$ operators 
implies the existence of joint distribution functions $P^{x,y}_{l}(\alpha,\beta)$, where 
$P^{x,y}_{l}(\alpha,\beta) d\alpha d\beta$ is the probability that the simultaneous measurements of 
$\hat{S}^{x}_{l}$ and $S^{y}_{l}$ give the values between $\alpha$ and $\alpha+d\alpha$,
and $\beta$ and $\beta+d\beta$, respectively. 
Here we provide the expression for $P^{x,y}_{l}(\alpha,\beta)$ and proves that this is indeed the 
unique solution. 

The joint distribution function $P^{x,y}_{l}(\alpha,\beta)$ is 
given by the following expression
\begin{flalign}
&P_{l}^{x,y}(\alpha,\beta) = \prod_{k} \int^{\pi}_{-\pi} \frac{d\lambda_{\theta sk}}{2\pi} \int^{\infty}_{0} \lambda_{rsk}e^{-\lambda_{rsk}^2/2} d\lambda_{rsk} \nonumber \\
&\times \delta\left(\alpha+i \beta - \rho \int^{l/2}_{-l/2} dr e^{i\chi( r,\{\lambda_{jsk} \}) } \right) 
\label{jointdist}
\end{flalign}
where the expression for $\chi( r,\{\lambda_{jsk} \})$ is given in Eq.(\ref{chi}).
To prove it, we first show that Eq.(\ref{jointdist}) reproduces the distribution 
function $P^{\theta}_{l}(\alpha)$ in Eq.(\ref{disttheta}) for all $\theta$. Then,
we show that a function with this property 
is unique, and therefore the expression in Eq.(\ref{disttheta}) 
is necessarily the joint distribution function. 

Given a joint distribution function $P^{x,y}_{l}(\alpha,\beta)$, we can determine
the distribution function $P^{\theta}_{l}(\gamma)$ of a spin pointing in the direction
$(\cos\theta, \sin\theta, 0)$. Consider the spin $\vec{S}$
in the $x-y$ plane with $\vec{S} = (\alpha, \beta, 0)$ whose probability distribution 
is given by $P^{x,y}_{l}(\alpha,\beta)$. 
The projection of the spin $\vec{S}$ onto the axis pointing in the direction 
$(\cos\theta, \sin\theta, 0)$ is given by $|S|\cos(\phi-\theta)$ where 
$|S| = \sqrt{\alpha^2 + \beta^2}$ is the magnitude of spin and 
$\phi$ is the angle Arg$(\alpha+i\beta)$.
After a simple algebra, we find $|S|\cos(\phi-\theta)
= \alpha \cos\theta + \beta \sin\theta$.
Then given a spin $\vec{S} = (\alpha, \beta, 0)$, if one measures
the spin along the direction $(\cos\theta, \sin\theta, 0)$, the measurement
result gives $\gamma$ if and only if $\gamma = \alpha \cos\theta + \beta \sin\theta$.
From this consideration, the probability distribution that the measurement 
along the direction $(\cos\theta, \sin\theta, 0)$ gives the value $\gamma$
is given by
\begin{equation}
P^{\theta}_{l}(\gamma) = \int d\alpha d\beta 
P^{x,y}_{l}(\alpha,\beta) \delta(\gamma -\alpha \cos\theta - \beta \sin\theta).
\label{jointcondition}
\end{equation}
Now, if we plug in the expression of Eq.(\ref{jointdist})
in Eq. (\ref{jointcondition}), we see that $P^{\theta}_{l}(\gamma)$ agrees with 
Eq. (\ref{disttheta}) for all $\theta$. 

Now we prove the uniqueness of a function with the above property, {\it i.e.}
a function that reproduces Eq. (\ref{disttheta}) through the relation Eq.(\ref{jointcondition}). 
Suppose you have 
another distribution $\tilde{P}^{x,y}_{l}(\alpha,\beta)$ 
that satisfies Eq.(\ref{jointcondition})
for all $\theta$. We define 
$Q(\alpha,\beta) = P^{x,y}_{l}(\alpha,\beta)-\tilde{P}^{x,y}_{l}(\alpha,\beta)$.
Our goal is to show that $Q(\alpha,\beta)$ must be equal to zero. 
By definition, we have the equality
\begin{equation}
0 = \int d\alpha d\beta 
Q(\alpha,\beta) \delta(\gamma -\alpha \cos\theta - \beta \sin\theta),
\label{qcondition}
\end{equation}
for all $\theta$ and $\gamma$. If we take the Fourier transform 
of both sides of Eq. (\ref{qcondition}) in terms of $\gamma$,  
we obtain
\begin{eqnarray*}
0 &=& \int d\gamma \int d\alpha d\beta 
Q(\alpha,\beta) \delta(\gamma -\alpha \cos\theta - \beta \sin\theta) e^{iw\gamma}\\
& = & \int d\vec{\alpha} Q(\vec{\alpha}) e^{i\vec{w}\cdot\vec{\alpha}}.
\end{eqnarray*}
In the last line, we defined $\vec{w} = w(\cos\theta , \sin\theta )$
and $\vec{\alpha} = (\alpha, \beta)$. Notice that this equation holds
for any $\vec{w}$. Then this last expression is just like (two-dimensional)
Fourier transform of $Q$. By taking the inverse fourier transform of the 
last expression in terms of $\vec{w}$, we find
\begin{eqnarray*}
0 &=& \int^{\infty}_{-\infty} d\vec{w} \int d\vec{\alpha}
Q(\vec{\alpha}) e^{i\vec{w}\cdot (\vec{\alpha}- \vec{\alpha'})}  Q(\vec{\alpha'}),
\end{eqnarray*}
thereby proving the uniqueness of 
the joint distribution $P^{x,y}_{l}(\alpha,\beta)$.

From the joint distribution function in Eq.(\ref{jointdist}), one can also obtain 
other distributions, such as the distribution $P^{\perp}_{l}(\gamma)$ of the square of the transverse spin magnitude, 
$\left(S^{\perp}_{l}(t) \right)^2$, which is given by
\begin{eqnarray}
P^{\perp}_{l}(\gamma) & =& \int^{\infty}_{-\infty}
 d\alpha d\beta P^{x,y}_{l}(\alpha,\beta) \delta\left(\gamma -\alpha^2 - \beta^2 \right) 
\nonumber \\
&=&\prod_{k} \int^{\pi}_{-\pi} \frac{d\lambda_{\theta sk}}{2\pi} \int^{\infty}_{0} \lambda_{rsk}e^{-\lambda_{rsk}^2/2} d\lambda_{rsk} \nonumber \\
&&\times \delta\left(\gamma - \left| \rho \int^{l/2}_{-l/2} dr e^{i\chi( r,\{\lambda_{jsk} \}) } \right|^{2} \right)  \label{magnitudedist}
\end{eqnarray}

 \begin{figure}[th]
\begin{center}
\includegraphics[width = 8cm]{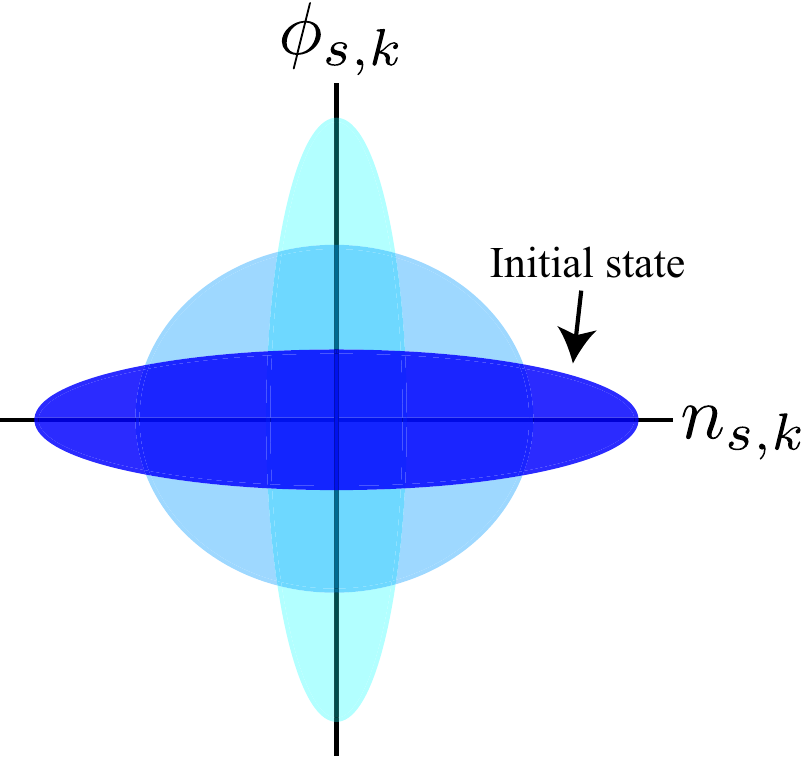}
\caption{The illustration of the dynamics for each harmonic oscillator mode, described by
the Hamiltonian Eq.(\ref{spinhamiltonian}). The initial state contains a large fluctuation of 
density difference $\hat{n}_{s,k}$ given by  $\braket{\hat{n}_{s,k} \hat{n}_{s,-k}} = \eta \rho/2$(see Eq.(\ref{spininistate})),
and its conjugate variable, the phase difference $\hat{\phi}_{s,k}$, has a small fluctuations. 
In the subsequent dynamics, such squeezed state evolves and energy oscillates between the fluctuations of 
the density difference and phase difference. }
\label{harmonicoscillator}
\end{center}
\end{figure}

\subsubsection{Interpretation of the distribution dynamics} \label{section:interpretation}
The form of the distribution function in Eq. (\ref{jointdist}) encapsulates the interpretation in terms of 
dynamics originating from 
spin waves explained in Sec~\ref{section:qualitative}. 
Here $e^{i\chi( r,\{t_{jsk} \})}$ represents the spin direction at coordinate $r$, 
where the $x-y$ plane of the spin component is taken to be a complex plane. 
Then Eq.(\ref{jointdist}) suggests that for a given instance of the set $\{\lambda_{jsk} \}$, 
$(S_{l}^{x}+ i S_{l}^{y})$ is simply the sum of the local spin directions $e^{i\chi( r,\{\lambda_{jsk} \})}$
over the integration length $l$. 
The local spin direction at position $r$ are determined by the phase $\chi( r,\{t_{jsk} \})$,
which receives contributions from each spin wave of momentum $k$ 
with strength $A_{k}(t) = \lambda_{rsk}\sqrt{\braket{|\hat{\phi}_{s,k}(t)|^2}}$. 
Spin waves with momenta $k$ rotate the spins as $\sin(kr+\lambda_{\theta s k})$
(see the expression of $\chi( r,\{\lambda_{jsk} \})$ in Eq.(\ref{chi})). 
The rotation strength $A_{k}(t) \propto \lambda_{rsk}$ has the distribution $\lambda_{rsk} e^{ - \lambda_{rsk}^2/2}$, which
represents the quantum fluctuation of the spins.
 On the other hand, $\lambda_{\theta sk}$ is distributed uniformly between $-\pi$ and $\pi$. 

  \begin{figure*}[th]
\begin{center}
\includegraphics[width = 17cm]{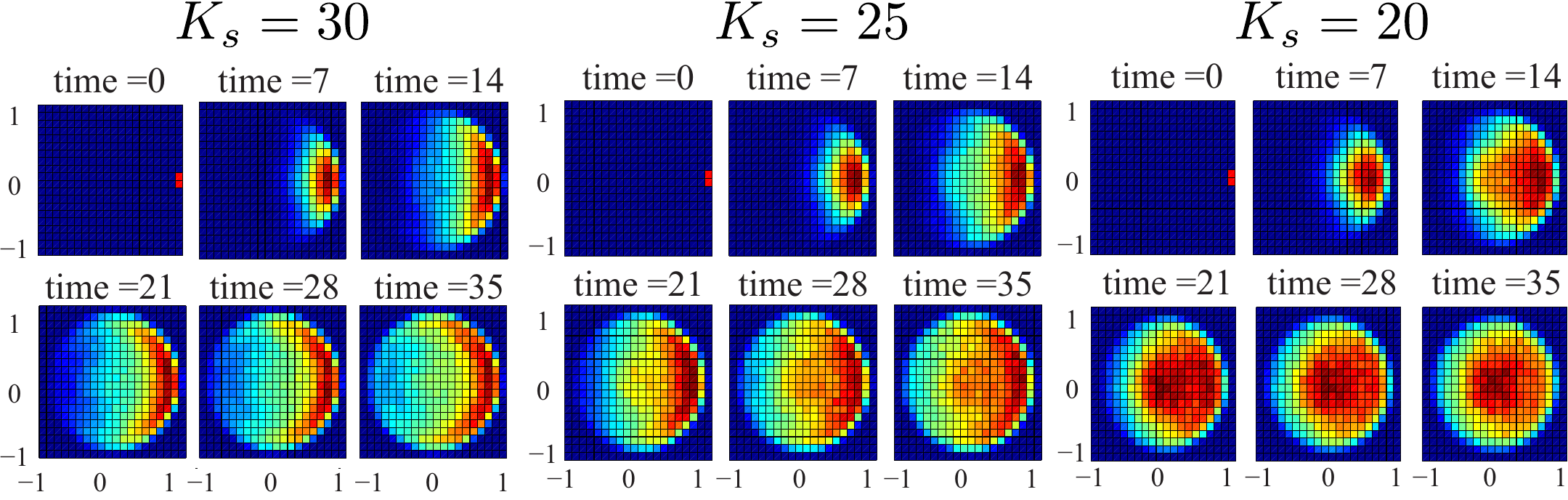}
\caption{The dynamics of the joint distributions for $L/\xi_{s} = 200$, $\xi_{s}=40$ and 
various spin Luttinger parameters $K_{s} = 30,25,$ and $20$. Here axes are scaled such that the maximum
value of $\alpha$ and $\beta$ are $1$. Smaller value of $K_{s}$ enhances
the spin fluctuations, leading to a stronger diffusion and decay. Time is measured in units of 
$\xi_{s}/c_{s}$.  }
\label{differentk}
\end{center}
\end{figure*}

The dynamics of phase fluctuations $\braket{|\hat{\phi}_{s,k}(t)|^2}$ can in fact be easily 
understood by considering the Hamiltonian given by Eq.(\ref{spinhamiltonian}) as a harmonic oscillator
for each $k$(Fig.~\ref{harmonicoscillator}). We first note that the initial state has a large fluctuation of density $\hat{n}_{s,k}$ because 
the initial $\pi/2$ pulse prepares the state in the (almost) eigenstate of $S^{x} = \rho \cos(\hat{\phi}_{s,k})$
with a small fluctuation of $\hat{\phi}_{s,k}$, the conjugate variable of $\hat{n}_{s,k}$.
The fluctuation of $\hat{n}_{s,k}$ is given by $\braket{\hat{n}_{s,k} \hat{n}_{s,-k}} = \eta \rho/2$
(see Eq.(\ref{spininistate})). 
 Because of this large fluctuation in  the density, almost all the energy of the initial state
 is stored in the interaction term $|n_{s,k}|^2$ in Eq.~(\ref{spinhamiltonian}). 
 Therefore, the total energy of each harmonic oscillator can be estimated as 
 $\frac{\pi c_{s} \rho \eta}{4 K_{s} }$.  During the dynamics dictated by the harmonic 
 oscillator Hamiltonian, 
 this energy oscillates between the density fluctuations and phase fluctuations in 
 a sinusoidal fashion, see Fig.~\ref{harmonicoscillator}.
In the dephased limit of the dynamics,  
approximately equal energy of the system is distributed to the phase and density 
 fluctuations, and from the conservation of energy, we conclude that
 the characteristic magnitude of phase fluctuation is given by $\braket{|\phi_{s,k}(t)|^2}\sim 
 \frac{ \pi^2 \rho \eta}{4 K^2_{s} k^2 }$. Such $1/k^2$ dependence of $\braket{|\hat{\phi}_{s,k}(t)|^2}$
 agrees with the more rigorous result in Eq.(\ref{kfluctuation}). 
 Therefore the spin fluctuations dominantly come from spin waves with long wavelengths, as we have 
 stated in Section \ref{section:qualitative}. Moreover, the weak dependence of spin dynamics on 
 high momenta contributions justifies the use of Tomonaga-Luttinger theory for describing the dynamics. 
 We will more carefully analyze the dependence of distributions on the high momentum cutoff in Sec~\ref{section:cutoff}.

 From the simple argument above, it is also clear that the spin fluctuations coming from
 spin waves with momenta $k$ have the time scales associated with the harmonic
 oscillators given by $\frac{1}{|k| c_{s}}$.
 Again, this rough argument agrees with the more rigorous result presented in 
 Eq.(\ref{kfluctuation}). 
 Therefore, the fast dynamics is dominated by spin waves with high momenta and 
 slow dynamics is dominated by low momenta. These considerations lead to the illustrative
 picture of Fig.\ref{illustration}. Furthermore, this implies that the dynamics of the magnitude of 
 spin $\left( \hat{S}^{\perp}_{l} \right)^2$ reaches a steady state around the time $\frac{l}{4 c_{s}}$
 since spin waves with wavelength longer than $l$ do not affect the magnitude. This should be 
 contrasted with the evolution of the $x$ component of spin which, in principle, keeps evolving 
 until the time scale of $\sim \frac{L}{4 c_{s}}$(see Fig.\ref{fig:jointdist}).

 The strength of interactions and correlations are associated with Luttinger parameter, $K_{s}$. 
 $K_{s}$ influences the spin fluctuations $\braket{|\phi_{s,k}(t)|^2}$ at all wavelength, and 
$\braket{|\phi_{s,k}(t)|^2}$ depends on $K_{s}$ as  $1/K_{s}^2$ for a fixed density. 
 As is expected, in the limit of the weak interaction corresponding to large $K_{s} $, the amplitude of spin fluctuation
 decreases. In Fig. \ref{differentk}, we have plotted the time evolution of the joint distributions 
 for $L/\xi_{s}=200$, $l/\xi_{s} =40$, and $K_{s}=20, 25$ and $30$. For larger $K_{s}$, 
 we see that the spin fluctuations get quickly suppressed.

\subsection{Dynamics of the expectation value of the magnitude of spin 
$\braket{\left(\hat{S}^{\perp}_{l}(t) \right)^2}$ } \label{magnitude}
In order to illustrate the dynamics of the Ramsey sequence further, it is helpful to study
the dynamics of the expectation value of the squared transverse magnitude, given by $\braket{\left(\hat{S}^{\perp}_{l}(t) \right)^2}$. 

\begin{figure}[th]
\begin{center}
\includegraphics[width = 7cm]{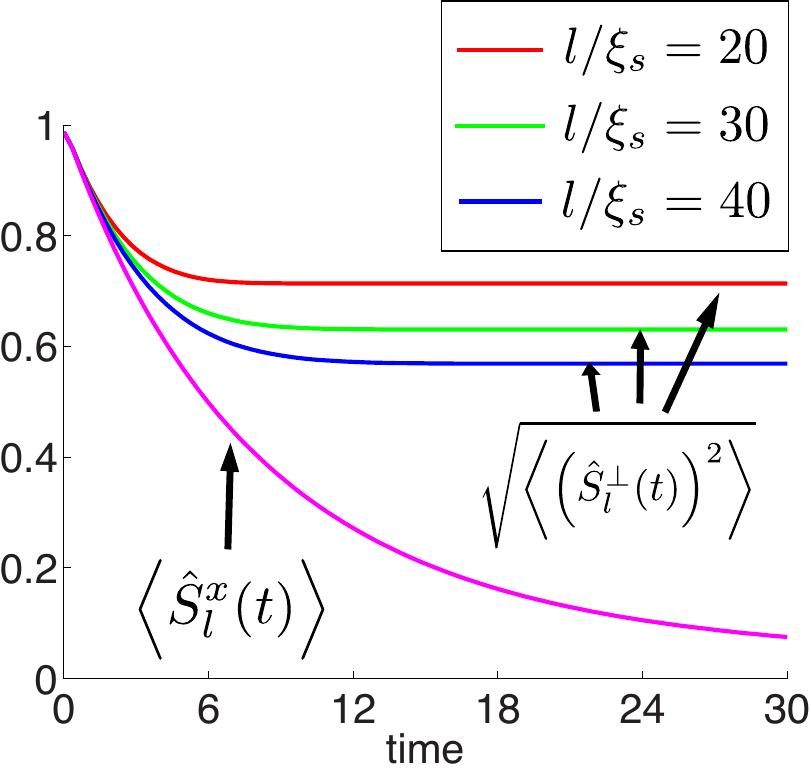}
\caption{The dynamics of the average value of the magnitude of spins,
 $\sqrt{\braket{\left(\hat{S}^{\perp}_{l}(t) \right)^2}}$, and the average of the $x$ component of spins $\braket{\hat{S}^{x}_{l}(t)}$. Here $y$ axis is scaled such that the initial values take the maximum value of $1$.
Here we took $L/\xi_{s} = 200$, $K_{s}=20$ and the integration lengths $l/\xi_{s} = 20,30,40$.
 The magnitude of spins decays only due to the spin waves with wavelengths shorter than the 
 integration length $l$, and the decay of the magnitude stops around the time scale of $\sim \frac{l}{4 c_{s}}$. 
 On the other hand, all spin waves contribute to the evolution of the of the $x$ component
of magnetization, which keeps decaying\cite{Bistritzer2007}.}
\label{average}
\end{center}
\end{figure}

In Fig.\ref{average}, we plot the evolution of $\sqrt{\braket{\left(\hat{S}^{\perp}_{l}(t) \right)^2}}$
with $K_{s}=20$, $L/\xi_{s} = 200$ and $l/\xi_{s} = 20,30,40 $. We also plotted 
$\braket{\hat{S}^{x}_{l}(t)}$ along with $\sqrt{\braket{\left(\hat{S}^{\perp}_{l}(t) \right)^2}}$ with the same parameters. 
It is easy to verify that $\braket{S^{x}_{l}(t)}$ is independent of integration length $l$\cite{Bistritzer2007}. 
As we have discussed in the previous section, $\sqrt{\braket{\left(\hat{S}^{\perp}_{l}(t) \right)^2}}$
reach the steady states at the time scale of $\frac{l}{4 c_{s}}$ with finite values, 
while $\braket{\hat{S}^{x}_{l}(t)}$ keeps decaying for much longer time.

It is interesting to ask if the long time limit of $\sqrt{\braket{\left(\hat{S}^{\perp}_{l}(t) \right)^2}}$ for 
sufficiently large integration length $l$ attains the value which corresponds to the 
one expected from the randomization of spin patches of size $\xi_{s}$. 
At low energies, spins within the length $\sim \xi_{s}$ 
are aligned in the same direction, but spin waves can randomize the direction of the spin 
for each of $l/\xi_{s}$ patches. Since the magnitude of spin within $\xi_{s}$ is $\xi_{s} \rho$, if the patches are 
completely randomized, the result of the random walk predicts that 
$\sqrt{\braket{\left(\hat{S}^{\perp}_{l} \right)^2}} \sim (\xi_{s} \rho)^{2} (l/\xi_{s})$. 
We will see below that, due to the properties of correlations in one dimension, 
the integrated magnitude of spin $\sqrt{\braket{\left(\hat{S}^{\perp}_{l} \right)^2}}$ 
never attains this form, albeit a similar expression is obtained (see Eq.(\ref{magnitude})).
Moreover, we identify the integration length $\tilde{l}$
which separates the "spin diffusion regime" and the "spin decay regime" by finding 
the decaying length scale for $\sqrt{\braket{\left(\hat{S}^{\perp}_{l} \right)^2}}$.

The results for the long time limit of $\braket{\left(\hat{S}^{\perp}_{l}(t) \right)^2}$ 
can be analytically computed. 
Following 
similar steps leading to Eq.(\ref{complicatedmoment}), we find 
\begin{eqnarray*}
 \braket{\left(\hat{S}^{\perp}_{l}(t) \right)^2} & = & \braket{ \left|\int dr \rho e^{i\phi(s,r)}\right|^2 } \\
& = & \prod_{i=1}^{2} \int^{l/2}_{-l/2}  \rho dr_{i}
\exp{ \left( -\frac{1}{2}\sum_{k \neq 0} \xi_{s,k}^{\{r_{i}\}} 
(\xi_{s,k}^{\{,r_{i}\}})^{*} \right)}.
\end{eqnarray*}
Here $\xi_{s,k}^{\{r_{i}\}} = |C_{s,k}|(e^{ikr_{1}} - e^{ikr_{2}})$.  
We introduce dimensionless variables $r'_{i} = r_{i}/l, k' = kl$ and 
the integration over $k$ in the exponent can be carried out as  
\begin{flalign*}
&\int dk' \frac{L}{2\pi l} \xi_{s,k'}^{\{r'_{i}\}} 
(\xi_{s,k'}^{\{,r'_{i}\}})^{*}  = \\ 
&\frac{2}{\pi} \int^{k'_{c}}_{k'_{min}} dk' \left( \frac{1}{\rho \eta l} \cos^2(|k|c_{s}t)
+ \frac{\pi^2 \rho \eta l }{k'^2 K_{s}^2} \sin^2(|k|c_{s}t)\right) \\
& \times \sin^2\left(\frac{r'_{1}-r'_{2}}{2}|k'|\right) \\
& \approx  \frac{k_{c}'}{2\rho \pi l \eta } + \frac{\pi \eta \rho l }{ 2K_{s} }|r'_{1}-r'_{2}|
\int^{\infty}_{0} dy \frac{ \sin^2(y)}{y^2}. 
\end{flalign*}
In the second line, we approximated $\cos^2(|k|c_{s}t) 
\approx \sin^2(|k|c_{s}t) \approx 1/2$, which is appropriate for long time. 
In the last line, we extended the upper limit of the integration for the second term 
to $\infty$ and the lower limit to $0$. The former is justified because we know that 
high momentum contribution is suppressed by $1/k^2$, and the latter is justfied because 
we also know low momenta excitations with wavelengths larger than $l$ do not affect $\hat{S}^{\perp}_{l}$. 
Since $\int^{\infty}_{0} dy \frac{ \sin^2(y)}{y^2} = \pi/2$, we find, in the long time limit,
\begin{flalign}
& \braket{\left( \hat{S}^{\perp}_{l} (t= \infty) \right)^2}
/\braket{\left( \hat{S}^{\perp}_{l} (t \approx 0) \right)^2} \nonumber \\ 
&=  2 \left\{ \frac{1}{l_{0}} - 
\left(\frac{1}{l_{0}}\right)^2 \left(1-\exp(-l_{0}) \right) \right\},   \label{magnitude}
\end{flalign}
where we expressed the result as a ratio of the asymptotic value and the value at shortest time scale 
of the theory given by $t \sim 1/\mu$. 
$l_{0} =\frac{\pi^2 \eta \rho l  }{8K_{s}^2 }$ is the dimensionless integration length
that controls the value of $\braket{\left( \hat{S}^{\perp}_{l} \right)^2}$ in the long time limit. 
As soon as $l_{0}$ becomes larger than $1$, the long time value of 
$\braket{\left(\hat{S}^{\perp}_{l} \right)^2}$ quickly 
approach the long integration limit, $\propto 2 \left\{ \frac{1}{l_{0}} - 
\left(\frac{1}{l_{0}}\right)^2 \right\}$. Therefore, $l_{0} \approx 1$ separates the "spin diffusion regime"
and the "spin decay regime."

An intuition behind the expression for $l_{0}$ can be explained through
the following heuristic argument. 
The system enters the spin decay regime when the spins within the integration length $l$ 
rotates by $2\pi$ across $l$. The angle difference between the spins at $r=0$ and $r=l$ 
in the long time limit
is roughly given by $\Delta \chi = \frac{1}{\sqrt{L}} \sum_{k} \lambda_{rsk} \sqrt{\braket{|\hat{\phi}_{s,k}|^{2}}_{mean}} \sin(kl)$
where $\braket{|\hat{\phi}_{s,k}|^{2}}_{mean}$ is the characteristic 
magnitude of $\braket{|\hat{\phi}_{s,k}(t)|^{2}}$ in Eq.(\ref{kfluctuation}), which is given by the half of the maximum magnitude of $\braket{|\hat{\phi}_{s,k}(t)|^{2}}$. 
Now the expectation of magnitude $\braket{(\Delta \chi)^{2}}$ over the quantum fluctuations represented by $ \lambda_{rsk}$ can be computed, and it yields 
$\braket{(\Delta \chi)^{2}} \approx \frac{\pi^2 \eta \rho l  }{4K_{s}^2 }$. 
When $\sqrt{\braket{(\Delta \chi)^{2}}}$ becomes of the order of $1$, the system enters the spin decay regime.
This estimate gives the boundary between the two regimes $l_{0} =\frac{\pi^2 \eta \rho l  }{8K_{s}^2 } \approx 1$ 
apart from an unimportant numerical factor.

It is notable that the Eq. (\ref{magnitude}) approaches
the random walk behavior $\propto (\xi_{s} \rho)^{2} (l/\xi_{s})$ very slowly, {\it i.e.}
in an algebraic fashion. Therefore, even in the steady state, the system retains 
a strong correlation among spins. Moreover, Eq. (\ref{magnitude}) in the limit of 
$l_{0} \rightarrow \infty$ is not just the random walk value, 
but is proportional to $K_{s}$, which measures the strength of fluctuations. 

The calculation above shows that the spin diffusion regime and the spin decay regime 
are separated at the integration length scale of $\tilde{l} \approx \frac{8K_{s}^2 }{ \pi^2 \eta \rho}$.
This length scale is nothing but the correlation length of spins in the long time limit. 
The calculation of the spin correlation length, for example, between $S^{x}(r)$ and 
 $\hat{S}^{x}(r')$ can be done similarly to the calculation of $\braket{\left(\hat{S}^{\perp}_{l} \right)^2}$.
 The result in the long time limit is 
 \begin{eqnarray}
 \braket{\hat{S}^{x}(r) \hat{S}^{x}(r')} \approx C \frac{\rho^2}{2} e^{- |r-r'|/\tilde{l}},
 \end{eqnarray}
 where $C = e^{-k_{c}/ (4 \pi \rho \eta)}$ is a small reduction of 
 the spins due to the contributions from high energy sector. 
 Thus, one expects qualitatively different behaviors of distribution functions 
 for integration lengths $l < \tilde{l}$ and $l > \tilde{l}$.

\subsection{Momentum cut-off dependence} \label{section:cutoff}
The description of dynamics presented above uses the low energy effective 
theory. In order to confirm the self-consistency of our approach, 
we check that the distributions of spins are not strongly affected by 
high energy physics, {\it i.e.} they weakly depend on high momentum cut-off.
We have seen an indication that this is indeed the case through
the weak fluctuations of phases for large $k$, 
$\braket{|\hat{\phi}_{s,k}|} \propto 1/k^2$, in Sec~\ref{section:interpretation}. 

First of all, we analyze the high momentum cut-off $k_{c} \sim 2\pi/\xi_{s}$ dependence of 
the average value of $\hat{S}^{x}_{l} $. From the discussion in Sec~\ref{section:nomixing},
it is straightforward to obtain that(here we ignore $k =0$ contribution)
\begin{flalign}
\braket{\hat{S}^{x}_{l}} =& 
\int^{l/2}_{-l/2} \frac{\rho}{2} dx \braket{e^{i\phi(x)} + e^{-i\phi(x)}} \nonumber \\
=&\rho l \exp \left(-\frac{1}{2}\sum_{k\neq0} |C_{s,k}|^2 \right),  \nonumber \\
\sum_{k \neq 0} |C_{s,k}|^2 =& \int^{k_{c}}_{-k_{c}} dk
\left(\frac{\cos^2(|k|c_{s}t)}{ 4\pi \rho \eta}+
\frac{\pi \rho \eta}{4 k^2 K_{s}^2} \sin^{2}(|k|c_{s}t) \right)  \nonumber \\
\approx & \frac{k_{c}}{2 \pi \rho \eta} + \frac{\rho c_{s}t \eta}{(2K_{s}/\pi )^2}, \label{cutoff}
\end{flalign}
where in the last line, we took the long time limit $t \gg \xi_{s}/c_{s}$ \cite{Bistritzer2007}. 
In this limit, only the first term in Eq.(\ref{cutoff}) depends on the cutoff $k_{c}$,
and moreover, the cutoff dependence is independent of time. 
The effect is to reduce the value of $\braket{\hat{S}^{x}_{l}}$ through the 
multiplication of a number close to one in the weakly interacting limit. 
For example, if we take $k_{c} = 2\pi/\xi_{c}$, then the cut-off dependent term
reduces the value by multiplying $\exp\left(- \frac{k_{c}}{4 \pi \rho \eta} \right) \approx e^{-1/(4K_{s})} \approx 1$.

In a similar fashion, higher moments of spin operators can be shown to have 
a weak dependence on the cutoff momentum $k_{c}$, as long as the integration
length is much larger than the healing length, $l/\xi_{s} \gg 1$. In this limit, 
$m$ moments of, for example, $\hat{S}^{x}_{l} $ is reduced by 
$\exp\left(- m\frac{k_{c}}{4 \pi \rho \eta} \right)$. Therefore, the full distribution function
is simply reduced by the multiplication of a number close to one 
$\exp\left(- \frac{k_{c}}{4 \pi \rho \eta} \right) \approx e^{-1/(4K_{s})} \approx 1$ 
in the weakly interacting regime. 
This gives the self-consistency check of our results in Sec~\ref{section:nomixing}

\section{Dynamics of Full Distribution Function in the presence of mixing between spin and charge degrees
of freedom} \label{section:mix}
In this section, we extend the analysis in Sec~\ref{section:nomix} to a more
general case, in which spin and charge degrees of freedom mix. 
We will see that the distribution functions even for this more general case 
have essentially the same structure as in Eq. (\ref{disttheta}), and are described 
by spin waves with fluctuations whose amplitude is determined 
by the fluctuations of phase $\braket{|\hat{\phi}_{s,k}|^2}$.
One important difference from the decoupling case is the dependence of
spin distributions on the initial temperature of the system. The thermal
excitations are present in the charge degrees of freedom in the initial state,
and such thermal fluctuations increase the value of $\braket{|\hat{\phi}_{s,k}|^2}$ 
through the coupling between spin and charge during the evolution.

\subsection{Hamiltonian and initial state}
In a generic system of two component bosons in one dimension,  
spin and charge degrees of freedom couple through the mixing Hamiltonian
in Eq.(\ref{mixinghamiltonian}). 
Yet, Hamiltonian in Eq.(\ref{hamiltonian}) is still quadratic and
it can be diagonalized. 
We define new operators $\hat{\phi}_{1}, \hat{\phi}_{2}, \hat{n}_{1}, \hat{n}_{2}$ by 
\begin{eqnarray}\label{diagonalize}
\left( \begin{array}{c} \hat{\phi}_{1} \\ \hat{\phi}_{2} \end{array}\right) 
 & = & 
 \left( \begin{array}{cc} \cos\kappa & \sin\kappa \\
 -\sin\kappa & \cos\kappa \end{array}\right) 
 \left( \begin{array}{c} \sqrt{s_{c}} \hat{\phi}_{c} \\ \hat{\phi}_{s} \end{array}\right),  \\
\left( \begin{array}{c} \hat{n}_{1} \\ \hat{n}_{2} \end{array}\right) 
 & = & 
 \left( \begin{array}{cc} \cos\kappa & \sin\kappa \\
 -\sin\kappa & \cos\kappa \end{array}\right) 
 \left( \begin{array}{c} \frac{1}{\sqrt{s_{c}}} \hat{n}_{c} \\ \hat{n}_{s} \end{array}\right).
\end{eqnarray}
Mixing angle $\kappa$ and scaling parameter $s_{c}$ are 
chosen so that the Hamiltonian is written in the following diagonal form
\begin{eqnarray}
H & = & H_{1} + H_{2}, \label{hamiltonianmixdiagonalized} \\
H_{1} &=& \int^{L/2}_{-L/2} dr \frac{\rho}{2m_{1}} (\nabla \hat{\phi}_{1}(r))^2 + g_{1} (\hat{n}_{1})^2, \nonumber \\
H_{2} &=& \int^{L/2}_{-L/2} dr \frac{\rho}{2m_{2}} (\nabla \hat{\phi}_{2})^2 + g_{2} (\hat{n}_{2})^2.
\nonumber
\end{eqnarray}
 Explicitly, $\kappa$ and $s_{c}$ are given by
\begin{eqnarray*}
s_{c} &=& \frac{\frac{g_{mix}\rho}{2m_{c}} + g_{s}g^{\phi}_{mix}}
{g_{c}g^{\phi}_{mix} + \frac{g_{mix}\rho}{2m_{s}}}, \quad 
\tan\kappa = \frac{-\kappa_{0} \pm \sqrt{\kappa_{0}^2+4}}{2}, \\
\kappa_{0} &=& \frac{s_{c} g_{c}-g_{s}}{\sqrt{s_{c}}g_{mix}}
 = \frac{\frac{\rho}{2m_{c}}
  - s_{c}\frac{\rho}{2m_{s}}}{g^{\phi}_{mix}\sqrt{s_{c}}},
 \end{eqnarray*}
 where $\pm$ in the expression of $\tan\kappa$ is $+$ when $\kappa_{0}>0$ and $-$ when $\kappa_{0}<0$. 
We defined $\kappa$ such that $\kappa =0$ corresponds to 
decoupling of charge and spin, i.e. to $g_{mix} =0$ and $g_{mix}^{\phi} = 0$ in Eq.(\ref{mixinghamiltonian}). 
Parameters $g_{1}$, $g_{2}$, $\frac{\rho}{2m_{1}}$ and
$\frac{\rho}{2m_{2}}$ are given by
\begin{eqnarray*} \label{gs}
g_{1} & = & s_{c} g_{c} + \sqrt{s_{c}} \tan\kappa g_{mix},  \\
g_{2}  &=&  g_{s} -\sqrt{s_{c}} \tan\kappa g_{mix}, \\
\frac{\rho}{2m_{1}} & = & \frac{\rho}{2m_{c} s_{c}} + \tan\kappa \frac{g^{\phi}_{mix} }{\sqrt{s_{c}}} ,\\
\frac{\rho}{2m_{2}} &=& \frac{\rho }{2m_{s}} - \tan\kappa \frac{ g^{\phi}_{mix}}{\sqrt{s_{c}}}.
\end{eqnarray*}
In the weakly interacting systems which we study in this paper, 
Luttinger parameters $K_{i}$ and sound velocities $c_{i}$ 
are determined for each Hamiltonian $H_{i}$, $i=\uparrow, \downarrow, c,s,1,2$ through 
\begin{equation}
K_{i} =\pi \sqrt{\frac{\rho}{2m_{i} g_{i}}}, \quad
c_{i} = \sqrt{\frac{\rho g_{i}}{m_{i} }}. \label{ki}
\end{equation}

At finite temperature, the state before the first $\pi/2$ pulse 
contains excitations, and these
excitations are carried over to the charge degrees of freedom
after the pulse. Pulse only acts on the spin
degrees of freedom, and the local sum density of spin-up and down is left untouched 
as long as the pulse is applied in a short time compared to the inverse of 
typical excitation energies, $\beta = 1/(k_{B}T)$. 
In other words, the local density fluctuation of spin-up, $\hat{n}_{\uparrow}(r)$, before $\pi/2$ pulse
is converted to the sum of the local density fluctuation of spin-up and spin-down,
$\hat{n}_{\uparrow}(r) + \hat{n}_{\downarrow}(r)$ after $\pi/2$ pulse. 
In this strong pulse limit, then, the distribution of $\hat{n}_{\uparrow}(r)$ before $\pi/2$ pulse
is the same as the distribution of $\hat{n}_{\uparrow}(r) + \hat{n}_{\downarrow}(r)$ after $\pi/2$ pulse. 

The distribution of the local density for spin-up atoms before $\pi/2$ pulse is determined by the
density matrix for spin-up given by $e^{-\beta H'_{\uparrow}}$, where in the weak interaction regime 
we have (see Eq.(\ref{hamiltonian}))
\begin{equation}
H'_{\uparrow} = \int^{L/2}_{-L/2} dr 
\left[ \frac{2\rho}{2m_{\uparrow}} (\nabla \hat{\phi}_{\uparrow}(r))^2 
+ g_{\uparrow\uparrow} (\hat{n}_{\uparrow}(r))^2 \right]. \nonumber 
\end{equation}
Then, the density matrix which produces the distribution of $\hat{n}_{\uparrow}(r) + \hat{n}_{\downarrow}(r)$
required above is given by  $e^{-\beta H_{c\uparrow}}$ where 
\begin{eqnarray}
H_{\uparrow c} = &\int^{L/2}_{-L/2} dr 
\left[ \frac{2\rho}{2m_{\uparrow}} \left\{ (\nabla \hat{\phi}_{\uparrow}(r)+\nabla \hat{\phi}_{\downarrow}(r))/2 \right\}^2 \right.,\nonumber   \\
& \left.+ g_{\uparrow\uparrow} (\hat{n}_{\uparrow}(r) + \hat{n}_{\downarrow}(r))^2 \right]\nonumber  \\
 = & \frac{c_{c\uparrow}}{2} \int^{L/2}_{-L/2} dr 
\left[\frac{K_{c\uparrow}}{\pi} (\nabla \hat{\phi}_{c})^2 
+\frac{\pi}{K_{c\uparrow}} \hat{n}_{c}^2 \right]. \label{charge_initial}
\end{eqnarray}
where $K_{c\uparrow} =\frac{\pi}{4} \sqrt{\frac{\rho}{m_{\uparrow} g_{\uparrow \uparrow}}}$ 
and $c_{c\uparrow} = \sqrt{\frac{2\rho g_{\uparrow \uparrow}}{m_{\uparrow} }} $.

The initial state for spins is determined by the $\pi/2$ pulse, and we obtained the state 
in Eq.(\ref{spininitial_state}). 
Then, the complete initial density matrix after the first $\pi/2$ pulse is given by
\begin{equation}
\hat{\rho}_{0} = \ket{\psi_{0}} \bra{\psi_{0}} 
\otimes e^{-\beta H_{c\uparrow}}/\textrm{Tr} \left(e^{-\beta H_{c\uparrow}} \right).
\label{initialdensity}
\end{equation}
This density matrix evolves in time as $\hat{\rho}(t) = e^{-itH} \hat{\rho}_{0}e^{itH}$. 
Since we assume that the preparation of the initial state is done through a
strong, short pulse, the spin and charge degrees of freedom 
are unentangled in the initial state. 

\subsection{Time evolutions of operators}
In order to calculate the distribution function of $\hat{S}^{\theta}_{l}$, we again
start from the calculation of $m$th moments, 
$\textrm{Tr}\left( \hat{\rho}(t) \left( \hat{S}^{\theta}_{l}\right)^m \right)$. Evaluation of moments 
can be done through a similar technique used in Sec~\ref{section:nomixing}.  

%
%

In the following, we describe convenient, time-dependent operators
$\gamma_{s,k}(t)$ and $\gamma_{c,k}(t)$ 
used to evaluate spin operators such as $e^{i \hat{\phi}_{s,k}}$. 
The first operator resides in the spin sector and it is again the annihilation operator 
of the initial spin state such that $\textrm{Tr} \gamma_{s,k}(0) \hat{\rho}_{0} = 0$. This operator 
is given in Eq. (\ref{gamma}), which is 
\begin{eqnarray*}
\gamma_{s,k}(t) & = & e^{-itH} \gamma_{s,k}(0) e^{itH}, \\
 \left( \begin{array}{c} \gamma_{s,-k}^{\dagger}(0) \\ \gamma_{s,k}(0) \end{array}\right) 
  &=&  
 \left( \begin{array}{cc} 
 \frac{1}{\sqrt{1-4|W_{k}|^2}} & \frac{-2W_{k}}{\sqrt{1-4|W_{k}|^2}}   \\ 
  \frac{-2W_{k}}{\sqrt{1-4|W_{k}|^2}} & \frac{1}{\sqrt{1-4|W_{k}|^2}}
  \end{array}\right) 
 \left( \begin{array}{c} b_{s,-k}^{\dagger} \\ b_{s,k} \end{array}\right), \label{gammask}
\end{eqnarray*}
with $2W_{k} =  \frac{1-\alpha_{k}}{1+\alpha_{k}}$ and 
$\alpha_{k} = \frac{|k|K_{s}}{\pi \rho \eta}$ as before. 
The second operator is the operator of charge degrees of freedom, and it is given by 
\begin{eqnarray*}
\gamma_{c,k}(t)  &=&  e^{-itH} \gamma_{c,k}(0) e^{itH}, \\
 \gamma_{c,k}(0) &=& b_{c \uparrow,k}.
\end{eqnarray*}
where $b_{c \uparrow,k}$ 
is an annihilation operator for the elementary excitations in $H_{c\uparrow}$. 
Since $\gamma_{s,k}(t)$ and $\gamma_{c,k}(t)$ commute at $t=0$, they commute 
at any time $t$. 
We will drop the time dependence of $\gamma_{a,k}(t)$ in the notation from now on.

From the expression of initial density matrix $\hat{\rho}_{0}$ in Eq. (\ref{initialdensity}), 
it is easy to check that the density matrix at time $t$ given by $\hat{\rho}(t) = e^{-itH}\hat{\rho}_{0}e^{itH}$
can be written as the tensor product of the density matrix of operators 
$\gamma_{s,k}(t)$ and that of $\gamma_{c,k}(t)$. This is because $\hat{\rho}_{0}$ 
is a tensor product of the density matrices of $\gamma_{s,k}(t=0)$ and that of $\gamma_{c,k}(t=0)$. 
This structure of the density matrices at time $t$ 
allows the independent evaluation of $\gamma_{s,k}(t)$ and  $\gamma_{c,k}(t)$ operators,
and it is advantageous to express spin operators in terms of these operators. 

As we show in the Appendix \ref{appendix:c}, we can write $\hat{\phi}_{s,k}$
in terms of $\gamma_{c,k}(t)$ and $\gamma_{s,k}(t)$ as follows. 
\begin{equation}
\frac{1}{\sqrt{L}} \hat{\phi}_{s,k} = C_{s,k}^{*} \gamma_{s,-k}^{\dagger} 
+ C_{s,k} \gamma_{s,k}
+ C_{c,k}^{*} \gamma_{c,-k}^{\dagger} + C_{c,k} \gamma_{c,k}, \label{cs}
\end{equation}
where explicit expression of $C_{s,k}$ and $C_{c,k}$ are given by 
\begin{widetext}
\begin{eqnarray} \label{explicitc}
C_{s,k} & = &  i \sqrt{\frac{1}{2L \rho \eta}}
\left( 
\left\{ \cos^2\theta \cos(c_{2}|k|t) 
+ \sin^2\theta \cos(c_{1}|k|t) \right\}  - i \frac{K_{s}}{\alpha_{k}} \left\{ \frac{\cos^2\theta \sin(c_{2}|k|t)}{K_{2}} 
+ \frac{\sin^2\theta \sin(c_{1}|k|t)}{K_{1}} \right\}
\right), \nonumber \\
C_{c,k} & = & \cos\theta \sin\theta \sqrt{\frac{\pi }{2L|k| s_{c} \tilde{K}_{c\uparrow}}}
\left( i \left\{\cos(c_{1}|k|t)- \cos(c_{2}|k|t) \right\} -\tilde{K}_{c\uparrow} \left\{ \frac{\sin(c_{2}|k|t)}{K_{2}} 
- \frac{\sin(c_{1}|k|t)}{K_{1}} \right\} \right), \label{cc}
\end{eqnarray}
\end{widetext}
where $\tilde{K}_{c\uparrow} = K_{c\uparrow}/\sqrt{s_{c}}$. 

Using Eq.(\ref{cs}), we find an expression for $(\hat{S}^{\theta}_{l})^m$ in terms of $\gamma_{a,k}$
with $a=s, c$ as follows
\begin{widetext}
\begin{eqnarray*}
(\hat{S}^{\theta}_{l})^m&=& 
\prod_{i=1}^{m} \int^{l/2}_{-l/2} dr_{i} \frac{\rho}{2} \sum_{\{s_{i}\}} 
e^{i\sum_{k\neq 0}(\xi_{s,k}^{*} 
\gamma_{s,k}^{\dagger} + \xi_{s,k} \gamma_{s,k})}
e^{i\sum_{k\neq 0}(\xi_{c,k}^{*} 
\gamma_{c,k}^{\dagger} + \xi_{c,k} \gamma_{c,k})} e^{ i(\sum_{i} s_{i})\phi_{s,0}/\sqrt{L} }e^{ -i(\sum_{i} s_{i})\theta },
\end{eqnarray*}
\end{widetext}
where $\xi_{a,k} = (\sum_{i}^{m} s_{i} e^{ir_{i}k}) C_{a,k}$. 
In the following, we separately 
evaluate three contributions; $k=0$ component given by $e^{ i(\sum_{i} s_{i})\phi_{s,0}/\sqrt{L} }$;
the charge component of $k \neq 0$ given by $e^{i\sum_{k\neq 0}(\xi_{c,k}^{*} 
\gamma_{c,k}^{\dagger} + \xi_{c,k} \gamma_{c,k})} $; 
the spin component of $k \neq 0$ given by $e^{i\sum_{k\neq 0}(\xi_{s,k}^{*} 
\gamma_{s,k}^{\dagger} + \xi_{s,k} \gamma_{s,k})}$.

\subsubsection{$k=0$ contribution}
The initial state of $k=0$ spin sector in Eq.(\ref{spininitial_state}) as well as that of the
charge sector in Eq. (\ref{charge_initial}) both have a Gaussian form
so that calculation of the trace 
Tr $\left\{e^{ i(\sum_{i} s_{i}) \phi_{s,0}/\sqrt{L} } \hat{\rho}(t) \right\}$ 
is straightforward. We leave the details to the Appendix \ref{appendix:k0}, and the result is 
\begin{flalign}
&\braket{e^{ i(\sum_{i} s_{i})\phi_{s,0}/\sqrt{L} }} 
= \exp\left( -\left(\sum_{i}s_{i}\right)^2 
\frac{\braket{\phi_{s,0}^2}_{t} }{2L} \right) \nonumber \\
&\braket{\phi_{s,0}^2}_{t} =
 \frac{1}{2\rho\eta } + \left(\sin^2\theta \frac{\pi c_{1}}{K_{1}} + 
\cos^2\theta  \frac{\pi c_{2}}{K_{2}}\right)^2  \frac{\rho\eta}{2}t^2 \nonumber \\
&+\sin^2\theta \cos^2\theta  
\left( \frac{\pi c_{1}}{K_{1}} - \frac{\pi c_{2}}{K_{2}} \right)^2 
\frac{\tilde{K}_{c\uparrow}}{\pi c_{c\uparrow} \beta }t^2.  \label{fluctuationmix0}
\end{flalign}

\subsubsection{$k \neq 0$, spin sector}
This calculation is analogous to Eq.(\ref{complicatedmoment}) and the result 
can be directly read off from Eq.(\ref{complicatedmoment}), and it is 
\begin{equation}
\braket{e^{i\sum_{k\neq 0}(\xi_{s,k}^{*} 
\gamma_{s,k}^{\dagger} + \xi_{s,k} \gamma_{s,k})}}
= \exp\left( -\frac{1}{2} \sum_{k \neq 0 } \xi_{s,k}^* \xi_{s,k} \right).
\label{spinpart}
\end{equation}

\subsubsection{$k \neq 0$, charge sector}
We first rewrite the density matrix at time $t$ as
\begin{eqnarray*}
\hat{\rho}_{c, k \neq 0}(t) &=& 
e^{-itH} e^{- \beta c_{c\uparrow} \sum_{k \neq 0} |k|b_{c \uparrow,k}^{\dagger}b_{c \uparrow,k}} 
e^{itH} /\mathcal{N} \\
& =&  e^{- \beta c_{c\uparrow} \sum_{k \neq 0} 
|k|\gamma_{c,k}^{\dagger}(t)\gamma_{c,k}(t)}  /\mathcal{N},
\end{eqnarray*}
where $\mathcal{N}_{c}$ is normalization given by
$\mathcal{N}_{c} = $ Tr$e^{- \beta c_{c\uparrow} \sum_{k \neq 0} 
|k|\gamma_{c,k}^{\dagger}(t)\gamma_{c,k}(t)} = \prod_{k \neq 0} -1/\lambda_{k}$ 
with $\lambda_{k} = e^{- \beta c_{c\uparrow}|k|}-1$. 

Then the trace of $\left(\hat{S}^{\theta}_{l}\right)^m$ for $k \neq 0$ spin
sector is 
\begin{flalign}
 &\braket{e^{i\sum_{k \neq 0} 
 (\xi_{c,k}^{*} \gamma_{c,k}^{\dagger} + \xi_{c,k} \gamma_{c,k})}} 
  = \\
&  \prod_{k \neq 0} \textrm{Tr}\left( e^{i
(\xi_{c,k}^{*} \gamma_{c,k}^{\dagger} + \xi_{c,k} \gamma_{c,k})} 
e^{-\beta |k| c_{c\uparrow} \gamma_{c,k}^{\dagger}\gamma_{c,k}}\right)/\mathcal{N} \nonumber
\end{flalign}
We can evaluate this by taking the trace in the basis of normalized coherent states $\ket{\alpha_{k}}$ 
such that $\gamma_{c,k}\ket{\alpha_{k}} = \alpha_{k} \ket{\alpha_{k}}$. 
The use of the identity $1 = \frac{1}{\pi} \int d^2 \alpha_{k}
 \ket{\alpha_{k}}\bra{\alpha_{k}}$ as well as of 
 an important equality $e^{v a^{\dagger}a} = :e^{(e^{v}-1)a^\dagger a}:$\cite{Imambekov2008a}, where $: \mathcal{O}:$
is a normal ordering of $\mathcal{O}$, leads to 
\begin{widetext}
\begin{flalign*}
 \braket{e^{i\sum_{k \neq 0}(\xi_{c,k}^{*} \gamma_{c,k}^{\dagger} + \xi_{c,k} \gamma_{c,k})}} 
 =& \frac{1}{\mathcal{N}}
\prod_{k \neq 0} \frac{1}{\pi} \int d^2 \alpha_{k} \bra{\alpha_{k}} e^{i
(\xi_{c,k}^{*} \gamma_{c,k}^{\dagger} + \xi_{c,k} \gamma_{c,k})} 
e^{-\beta |k| c_{c\uparrow} \gamma_{c,k}^{\dagger}\gamma_{c,k}} \ket{\alpha_{k}} \\
=& \frac{1}{\mathcal{N}} \prod_{k \neq 0}
\frac{1}{\pi} \int d^2 \alpha_{k} 
e^{-1/2 \xi_{c,k}^{*} \xi_{c,k}} \bra{\alpha_{k}} e^{i
\xi_{c,k}^{*} \alpha_{k}^{*}}  e^{i\xi_{c,k} \gamma_{c,k}} 
:e^{\lambda_{k} \gamma_{c,k}^{\dagger}\gamma_{c,k}}: \ket{\alpha_{k}} \\
 = & \prod_{k \neq 0}
e^{ -\frac{1}{2} \frac{1 + e^{-\beta |k| c_{c\uparrow}}}{1 - e^{-\beta |k| c_{c\uparrow}}}  \xi_{c,k}^{*} \xi_{c,k}}.
\end{flalign*}
\end{widetext}

\subsubsection{Full distribution function}

\begin{figure*}[th]
\begin{center}
\includegraphics[width = 13cm]{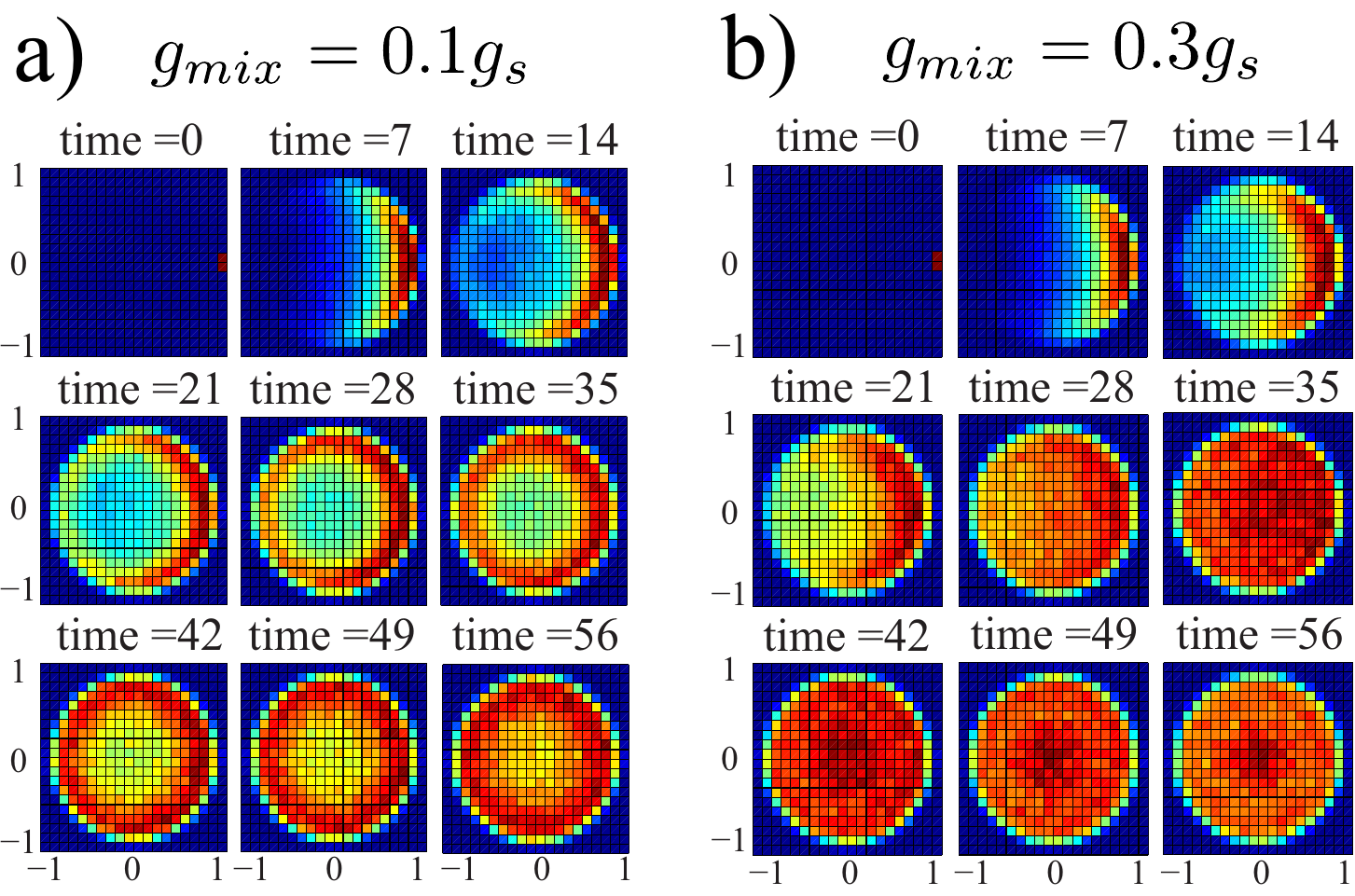}
\caption{ Time evolution of the joint distribution function $P^{x,y}(\alpha,\beta)$
for the system size $L/\xi_{s} =400$, the spin Luttinger parameter $K_{s} =20$ and 
integration length $l//\xi_{s}= 20$ in the presence of mixing between the spin and charge modes. 
For a), the interaction strength ratio is taken to be $g_{c}: g_{s}: g_{mix} = 1: 1 : 0.1$,
and for b), $g_{c}: g_{s}: g_{mix} = 1: 1 : 0.3$. Time is measured in units of 
$\xi_{s}/c_{s}$ where $c_{s}$ is the spin sound wave velocity. Here axes are scaled such that the maximum
value of $\alpha$ and $\beta$ are $1$.
With increasing strength of 
mixing, the large initial temperature affects the spin dynamics at earlier time more strongly. }
\label{figure:mix}
\end{center}
\end{figure*}

We can summarize the results above as 
\begin{widetext}
\begin{eqnarray}
\braket{ (\hat{S}^{\theta}_{l})^n } =
\sum_{\{s_{i}\}}  \prod_{i=1}^{m} \int dr_{i} \frac{\rho}{2} 
\exp\left( -\frac{1}{2} \sum_{k \neq 0 } \xi_{s,k}^* \xi_{s,k} \right) 
\exp\left( -\frac{1}{2} \sum_{k \neq 0 }M_{c,k} \xi_{c,k}^* \xi_{c,k} \right) 
\exp\left( -\frac{1}{2} \left(\sum_{i} s_{i} \right)^2 
\frac{\braket{\phi_{s,0}^2}_{t}}{L} \right) 
e^{ -i(\sum_{i} s_{i})\theta } \nonumber\\ \label{fullnthmoments}
\end{eqnarray}
\end{widetext}
Here, $M_{c,k} = \frac{1 + e^{-\beta |k| c_{c\uparrow}}}{1 - e^{-\beta |k| c_{c\uparrow}}}$.
As before, we introduce the auxiliary variables to separate spatial integrations over $r_{i}$. 
We can combine
$ \xi_{s,k}^* \xi_{s,k}$ and $M_{c,k} \xi_{c,k}^* \xi_{c,k}$ so that
we only need to introduce three sets of variables, 
$\lambda_{1,s,k}, \lambda_{2,s,k}, \lambda_{0}$ for Hubbard-Stratonovich transformation. 
Summing over $\{ s_{i} \}$ simplifies the result, leading to the following expression for the full distribution function
\begin{eqnarray}
P^{\theta}_{l}(\alpha) = &\prod_{k} \int^{\pi}_{-\pi} \frac{d\lambda_{\theta sk}}{2\pi} \int^{\infty}_{0} \lambda_{rsk} e^{-\lambda_{rsk}^2/2} d\lambda_{rsk}, \nonumber \\
\times& \delta\left(\alpha - \rho \int^{l/2}_{-l/2} dr \cos\left[\chi( r,\{\lambda_{jsk} \})-\theta \right] \right) \nonumber \\
\chi(r, \{\lambda_{jsk} \})  =& \sum_{k} \sqrt{\frac{\braket{|\hat{\phi}_{s,k}|^2}} {L}}
 \lambda_{rsk} \sin(kr +\lambda_{\theta sk} ),
\nonumber \\
\braket{|\hat{\phi}_{s,k}|^2} /L=&  |C_{s,k}|^2 + \frac{1 + e^{-\beta |k| c_{c\uparrow}}}{1 - e^{-\beta |k| c_{c\uparrow}}} |C_{c,k}|^2, \quad \textrm{$k \neq 0$}. \label{fluctuationmix} 
\end{eqnarray}
The last line can be confirmed by directly computing $\braket{|\hat{\phi}_{s,k}|^2}$,
using the expression in Eq. (\ref{cs}). The expression for $\braket{|\phi_{s,0}|^2}$ 
is given by Eq. (\ref{fluctuationmix0}). 
As before, the joint distributions as well as the distributions of squared transverse magnitude can
be obtained through the same procedure as in Sec~\ref{section:jointdist}.

The spin distribution in the presence of mixing between spin and charge degrees of 
freedom resembles the one in the absence of such mixing, and the only change is 
the additional contributions to phase fluctuations coming from the thermal excitations,
represented by $\frac{1 + e^{-\beta |k| c_{c\uparrow}}}{1 - e^{-\beta |k| c_{c\uparrow}}} |C_{c,k}|^2$ 
in Eq. (\ref{fluctuationmix}). 
$|C_{c,k}|$ is proportional to $\sin^{2}\kappa$ as one can see from Eq. (\ref{cc}). Thus, for 
weak coupling of $\kappa \sim 0$, the contribution is diminished by a factor of $\kappa^2$. 

In the experiment by Widera {\it et al.}, they used Rb$^{87}$ in the presence of 
Feshbach resonance. They employed the theory which assumes the absence of mixing 
between spin and charge degrees of freedom to analyze the decay of the Ramsey fringes. 
The ratio of interaction strengths in their experiment can be roughly estimated 
as $g_{c}:g_{s}:g_{mix} \approx 3.66:0.34:0.06$ which leads to the value of 
$\kappa \approx 2 \times 10^{-2}$. Therefore, the thermal contributions
are diminished by about four order of magnitude and thus, their assumption of decoupling
between spin and charge is justified.

In Fig.\ref{figure:mix}, we have plotted the evolution of the joint distribution functions for different 
strength of the coupling $g_{mix}$ at a relatively large initial temperature 
$k_{B} T = 0.4 \times 2\pi c_{c\uparrow}/\xi_{s}$ where $2\pi c_{c\uparrow}/\xi_{s}$ is approximately 
the high energy cut-off of Tomonaga-Luttinger theory. Here we took the system size
$L/\xi_{s} =400$, the Luttinger parameter $K_{s} = 20$, integration length $l= 20 \xi_{s}$. 
For Fig.\ref{figure:mix} a), the ratio of interaction is taken to be $g_{c}: g_{s}: g_{mix} = 1: 1 : 0.1$,
and for For Fig.\ref{figure:mix} b), $g_{c}: g_{s}: g_{mix} = 1: 1 : 0.3$. One can see that with increasing strength of 
mixing, the large initial temperature affects the spin dynamics at earlier time more strongly.  
For comparison, also see Fig.~\ref{fig:jointdist}.

\section{Interference of two one-dimensional condensates} \label{section:interference}
\subsection{Dynamics of interference pattern}
As we have described in Sec.\ref{summary:interference}, 
the full distribution of interference patterns can be studied in exactly the same 
way as we have studied the full distribution of spins in previous sections. 
In the following, we more formally describe the dynamics of split condensates. 

The low energy effective Hamiltonian of two quasi-condensates after splitting is given by 
\begin{eqnarray}
H & = & H_{L} + H_{R},  \label{splittedhamiltonian} \\
H_{L} &=& \int^{L/2}_{-L/2} dr 
\left[ \frac{\rho_{L}}{2m} (\nabla \hat{\phi}_{L}(r))^2 
+ \frac{g}{2} (\hat{n}_{L}(r))^2 \right], \nonumber \\
H_{R} &=& \int^{L/2}_{-L/2} dr 
\left[ \frac{\rho_{R}}{2m} (\nabla \hat{\phi}_{R}(r))^2 
+ \frac{g}{2} (\hat{n}_{R}(r))^2 \right]. \nonumber 
\end{eqnarray}
where we assumed weakly interacting bosons with a possible density difference 
$\rho_{R}- \rho_{L} \neq 0$ between the two condensates.
Here and in the following, we consider the rotating frame and ignore the chemical potential difference
$g/2(\rho_{R}^2- \rho_{L}^2)$ between left and right condensates arising from interactions. 

The interference pattern measures the phase difference $\hat{\phi}_{L}- \hat{\phi}_{R}$.
We describe the system in terms of the "spin" variables that
are the difference of left and right condensates and "charge" variables that are the sum 
of the two. Using the variables $\hat{\phi}_s = \hat{\phi}_R - \hat{\phi}_L$, $\hat{\phi}_c = \hat{\phi}_R + \hat{\phi}_L$,
$\hat{n}_s = (\hat{n}_R - \hat{n}_L)/2$, $\hat{n}_c = (\hat{n}_R + \hat{n}_L)/2$ , we find the Hamiltonian of the system to be
\begin{eqnarray}
H & = & H_{s} + H{c} + H_{int} \\
H_{s} &=& \int dx \left[ \frac{\rho_R + \rho_L}{8m} (\partial_x \hat{\phi}_{s})^2 + g \hat{n}_{s}^2 \right], \label{phasedifference} \\
H_{c} &=& \int dx \left[\frac{\rho_R + \rho_L}{8m} (\partial_x \hat{\phi}_{c})^2 + g \hat{n}_{c}^2 \right], \label{phasesum} \\
H_{int} &=& \int dx \left[\frac{\rho_R - \rho_L}{4m} \partial_x \hat{\phi}_{c} \partial_x \hat{\phi}_{s} \right]. \label{interaction}
\end{eqnarray}
Therefore, when the splitting makes two identical quasi-condensates with equal density, 
"spin" and "charge" degrees of freedom decouple and we can use a simpler theory 
derived in Sec~\ref{section:nomixing}. On the other hand, when the splitting makes two condensates
with unequal densities, more general theory of Sec~\ref{section:mix} needs to be employed.
In any case, the full time evolution of the distributions of interference patterns can be obtained,
which in principle can be compared with experiments. 

It is notable that the mixing of the "spin" and "charge" degrees of freedom 
for small density difference $\rho_{R}- \rho_{L} $ is not "small," in the sense
that the mixing angle $\kappa$ defined in section \ref{section:mix} takes the 
maximum value $\pi/4$. The spin decoupling in the limit of $\rho_{R}- \rho_{L} \rightarrow 0$
is recovered not by taking $\kappa \rightarrow 0$, but rather, by taking 
the time at which the effect of the coupling takes place to infinity. This is most explicitly 
shown in Eq.~(\ref{cc}) where the charge contributions of 
fluctuations go to zero as $c_{1} \rightarrow c_{2}$ which is attained in the limit 
$\rho_{R}- \rho_{L} =0$. 

\subsection{Interference patterns in equilibrium}
The techniques to calculate the full distribution functions presented 
in previous sections are directly applicable to also obtaining a simple form of the full distribution
functions of the interference patterns between two independent, thermal quasi-condensates. 
This problem has been previously analyzed in theory\cite{Gritsev2006,Stimming2010} as well as
in experiments\cite{Hofferberth2008,Betz2011}.

We consider the preparation of two independent one dimensional quasi-condensates.
If they are prepared by cooling two independent quasi-condensates, the temperature 
of the left quasi-condensate $T_{L}$ and 
that of right quasi-condensate $T_{R}$ are generically different. The density matrix of the initial state
is described by $\hat{\rho}_{0} = e^{-(\beta_{L} H_{L}+ \beta_{R} H_{R})}$ where 
$\beta_{a} = 1/(k_{B} T_{a})$ with $a=L, R$.
It is important to note that the constant shift of phase $\phi_{a} \rightarrow \phi_{a}
+ \theta_{ac}$ does not change the energy of the system, 
so that for the average over thermal ensemble one has to 
integrate over $\theta_{ac}$. Physically, this simply means that the phases of 
independent condensates are random. Then the only interesting 
distribution here is the distribution of the interference contrast
\cite{Polkovnikov2006,Hofferberth2008, Imambekov2008a, Gritsev2006} given by, 
\begin{equation}
\hat{C}^2 = \left| \int^{l/2}_{-l/2} e^{-i \hat{\phi}_{s}(r)} \right|^2 dr \label{contrast}
\end{equation}
which corresponds to,
in spin language, the squared transverse magnitude of the spin $\left( \hat{S}^{\perp}_{l} \right)^2$. 
The analysis of the evaluation of distributions in the density matrix of thermal 
equilibrium state in Sec~\ref{section:mix} can be directly extended to this case,
and we obtain the distribution 
 \begin{flalign}
P^{\perp}_{l}(\gamma) =&\prod_{k} \int^{\pi}_{-\pi} \frac{d\lambda_{\theta sk}}{2\pi} \int^{\infty}_{0} \lambda_{rsk} e^{-\lambda_{rsk}^2/2} d\lambda_{rsk} \nonumber \\
&\times \delta\left(\gamma - \left| \rho \int^{l/2}_{-l/2} dr e^{i\chi( r,\{\lambda_{jsk} \}) } \right|^{2} \right) 
\label{distthermal}, \\
\chi(r, \{\lambda_{jsk} \})  =& \sum_{k} \sqrt{\frac{\braket{|\hat{\phi}_{s,k}|^2}} {L}}
 \lambda_{rsk} \sin(kr +\lambda_{\theta sk} ), \nonumber \\
\braket{|\hat{\phi}_{s,k}|^2}  = & 
\frac{1+e^{-\beta_{L} |k|c_{L}}}{1-e^{-\beta_{L} |k|c_{L}}} \frac{\pi}{2|k|K_{L}} 
\nonumber \\ &+ \frac{1+e^{-\beta_{R} |k|c_{R}}}{1-e^{-\beta_{R} |k|c_{R}}} \frac{\pi }{2|k|K_{R}}, \label{thermalkfluctuation}
\end{flalign}  
where $c_{a}$ and $K_{a}, a=L,R$ are the sound velocity and Luttinger parameters 
of left and right quasi-condensate.

\subsection{Prethermalization of interference patterns} \label{section:prethermalization}
In Sec.\ref{summary:prethermalization}, we gave a heuristic argument for prethremalization phenomena,
where the distribution of the interference contrast amplitudes of the two {\it non-equilibrium} quasi-condensates 
are given by that of two {\it equilibrium} quasi-condensates at some effective temperature $T_{\textrm{eff}}$.  
We identified the effective temperature to be the energy stored in each momentum mode. In 
Sec.~\ref{section:interpretation}, we found this energy to be $\frac{\pi c_{s} \rho \eta}{4 K_{s} } $, thus we conclude 
$k_{B} T_{\textrm{eff}}= \frac{\pi c_{s} \rho \eta}{4 K_{s} } $. 

In the following, we formally derive the result above, using the expressions of full distributions of
interference patterns. 
The distribution of the interference contrast is determined by 
 $\braket{|\hat{\phi}_{s,k}|^2}$ given in Eq.(\ref{kfluctuation}). In the long time limit,
 we can take $\sin^2(c_{s}|k|t) \sim \cos^2(c_{s}|k|t) \sim 1/2$. 
 Moreover, since the interference contrast is most affected by the excitations with 
 small wave vectors $k$ with $\alpha_{k} = \frac{|k|K_{s}}{\pi \rho \eta} <1$, we can approximate
  the expression as 
 \begin{equation}
 \braket{|\hat{\phi}_{s,k}|^2} \approx  \frac{\pi}{2|k|K_{s}} \frac{\pi \rho \eta}{2|k|K_{s}}. \label{longtimefluctuation}
 \end{equation}
 
 On the other hand, for two quasi-condensates in thermal equilibrium, the 
 position of the interference peaks is again random. The interference contrast is determined by 
 $\braket{|\hat{\phi}_{s,k}|^2}$ given in Eq.(\ref{thermalkfluctuation}). Since 
 the main contribution to the fluctuation comes from low momenta, we approximate
 $e^{-\beta |k|c} \approx 1-  \beta |k|c$. It is easy to check that the sound velocity and 
 Luttinger parameters for each condensate is related to those of the difference mode 
 (see Eqs.(\ref{phasedifference}-\ref{interaction})) as $c_{L} = c_{R} = c_{s}$ and $K_{L} = K_{R} = 2 K_{s}$. 
Thus we obtain
 \begin{equation}
 \braket{|\hat{\phi}_{s,k}|^2}  \approx
\frac{2}{\beta |k|c_{s}} \frac{\pi}{2|k|K_{s}}. \label{hightempfluctuation}
\end{equation}
 
 Now the crucial observation is that our closed form expressions for distributions 
 of interference contrasts of both split quasi-condensates and thermal quasi-condensates
 are determined solely by  $\braket{|\hat{\phi}_{s,k}|^2}$, and they take precisely the same form 
 in terms of $\braket{|\hat{\phi}_{s,k}|^2}$. Moreover, the expressions given by
 Eq.(\ref{longtimefluctuation}) and Eq.(\ref{hightempfluctuation}) have the same dependence on wave vectors $|k|$.
 Therefore, the {\it full} distribution of interference contrast of split condensates become indistinguishable from
 that of thermal condensates with temperature 
 \begin{eqnarray}
 k_{B} T_{\textrm{eff}} &\approx&  \frac{\pi c_{s} \rho \eta}{4 K_{s} } = \frac{\mu \eta}{2},  \label{effectivetemp}
 \end{eqnarray} 
 where the second equality holds for weakly interacting bosons and the chemical potential 
 of one quasi-condesate is given by $\mu=g \rho$. 
 Thus, split one dimensional quasi-condensates indeed display the prethermalization phenomenon. 
 
\begin{figure}[t]
\begin{center}
\includegraphics[width = 8.5cm]{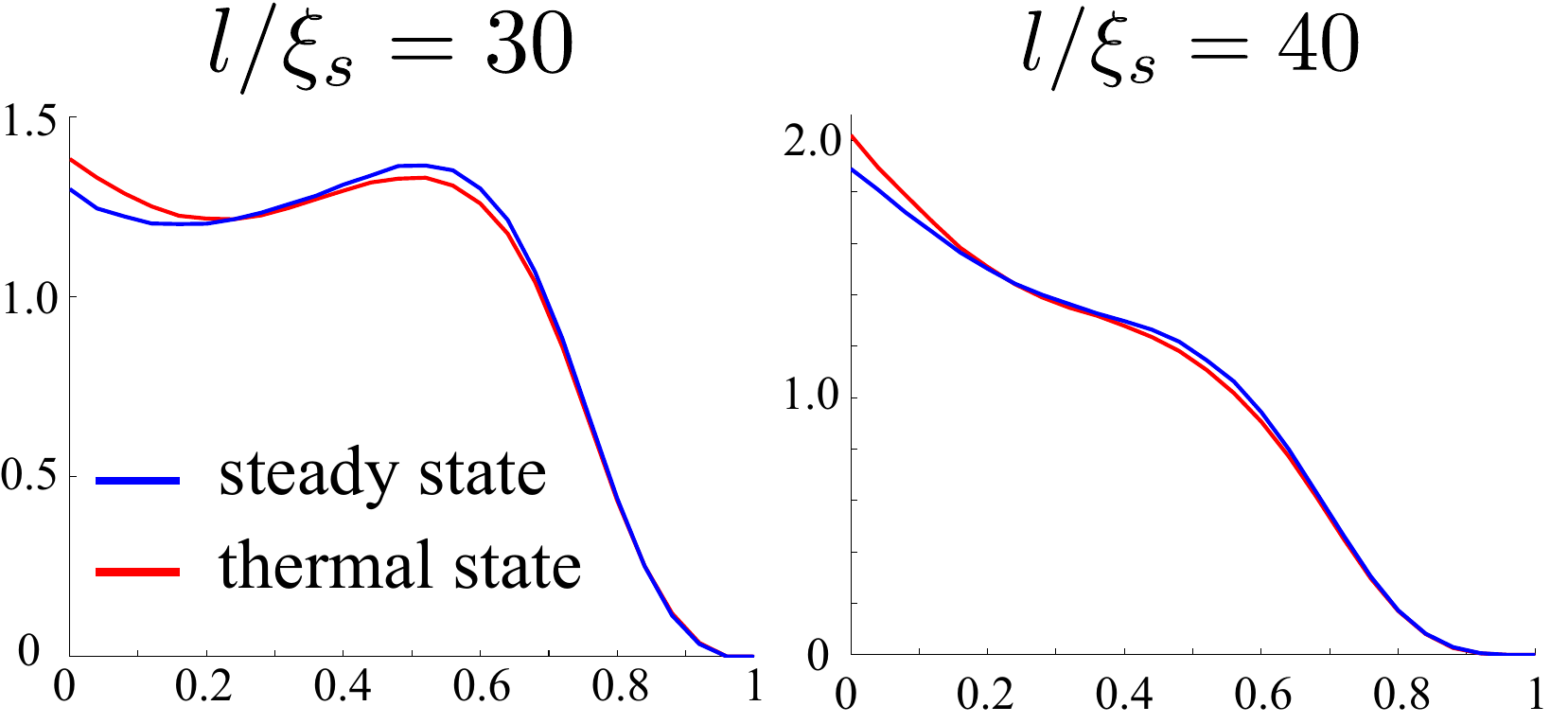}
\caption{ The distributions of interference contrast 
for steady states of split quasi-condensates and two thermal quasi-condensates.
Here $x$ axis is scaled such that the maximum
value of interference contrast is $1$.
For the split condensates, we plot the distribution at time 
$t = 60 \xi_{s}/c_{s}$ for Luttinger parameter $K_{s} =20$, system size $L=400 \xi_{s}$
 and two different integration length $l/\xi_{s} = 30, 40$. The thermal quasi-condensates 
 are for temperature $\frac{\pi c_{s}}{2 \xi_{s}}$ for the same integration length 
 corresponding to the effective temperature 
 obtained in Eq. (\ref{effectivetemp}). }
\label{figure6}
\end{center}
\end{figure}
 
 In Fig.~\ref{figure6}, we plot the interference contrast $P_{l}^{\perp}(\gamma)$ (see Eq.(\ref{magnitudedist}))
 of split condensates in a steady state 
 at time $t = 60 \xi_{s}/c_{s}$ for Luttinger parameter $K_{s} =20$, system size $L=400 \xi_{s}$
 and two different integration lengths $l/\xi_{s} = 30, 40$. Also we plot the interference contrast of the thermal 
 quasi-condensates (see Eq.(\ref{distthermal})) at temperature $\frac{\pi c_{s}}{2 \xi_{s}}$ for the same integration length.
 This temperature corresponds to the effective temperature 
 obtained in Eq. (\ref{effectivetemp}). Indeed we see only a small difference between the distributions of 
 steady states and thermal states for both integration lengths. The small difference comes from 
 the approximations made in obtaining the expressions  given by Eq.(\ref{longtimefluctuation}) and Eq.(\ref{hightempfluctuation}).

 
In the previous paragraphs, we assumed that the splitting prepares quasi-condensates with identical average
densities. Here we briefly consider the case in which 
the splitting process prepares two quasi-condensates with slightly different 
densities. In this case, the temperature of the initial quasi-condensates affects the interference 
contrast around the time scale of $\frac{\xi_{s}}{ (c_{L}  -c_{R}) \pi} 
\approx \frac{\hbar}{\mu (\sqrt{\rho_{L}}-  \sqrt{\rho_{R}})}$,
whereas the prethermalized, long-time transient state is reached around $\frac{\hbar}{\mu} \frac{l}{\xi_{s}}$, where $l$ is the 
integration length.

\begin{figure}[th]
\begin{center}
\includegraphics[width = 8cm]{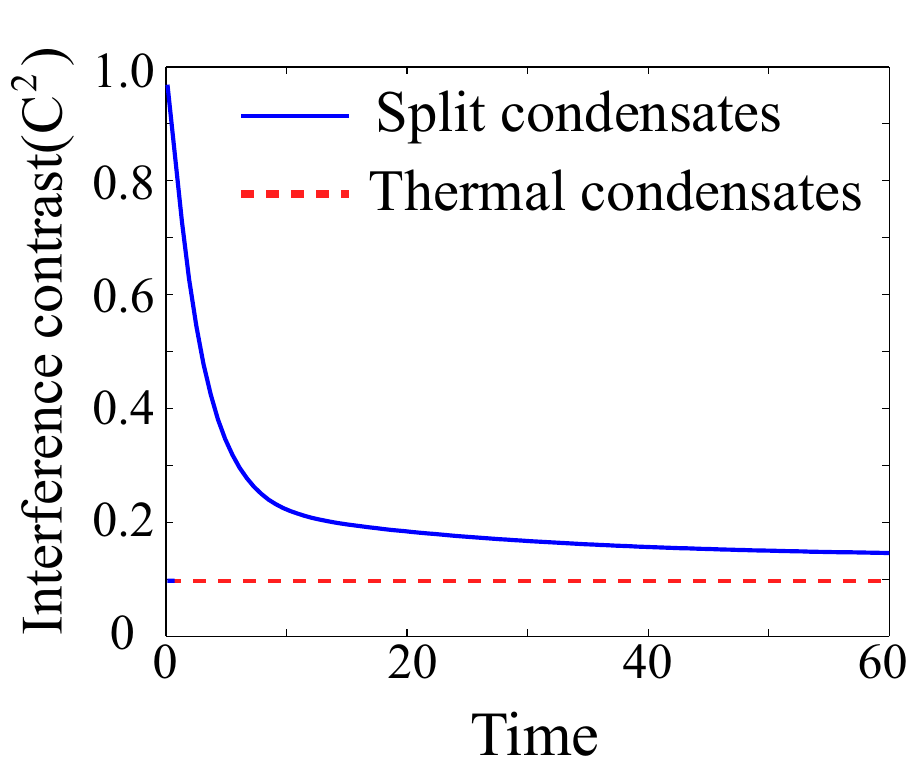}
\caption{ The evolution of the interference contrast $\hat{C}^2$ for system size $L= 500 \xi_{s}$, integration length $l = 40 \xi_{s}$, and the effective spin Luttinger parameter $K_{s} =20$ with initial temperature corresponding to the chemical potential $\mu$.  
Time is measured in units of $\xi_{s}/c_{s}$.
Here we took the density $\rho_{R} = 1.2 \rho$ and $\rho_{L} =0.8 \rho$. 
The magnitude of $\hat{C}^2$ for two thermal quasi-condensates at temperature $k_{B}T = \mu$ 
is plotted as red dotted line for comparison.}
\label{figure7}
\end{center}
\end{figure}

These analytic arguments can be confirmed through numerical simulations. 
In Fig.~\ref{figure7}, we have plotted the evolution of the interference contrast $\hat{C}^2$ 
for system size $L= 500 \xi_{s}$, integration length $l = 40 \xi_{s}$, and Luttinger parameter
$K_{s} =20$ with initial temperature corresponding to the chemical potential $\mu$. 
Here we consider a situation where the density of left quasi-condensate is 
different from that of right quasi-condensate by 20\% such that $\rho_{R} = 1.2 \rho$ and
$\rho_{L} = 0.8 \rho$ where $\rho_{R(L)}$ is the average density of the right (left) condensate,
and $2\rho$ is the average density of the initial condensate before splitting. Also for comparison we have plotted the 
magnitude of interference contrast of two quasi-condensates in {\it thermal} states 
at temperature given by $\mu$. From the plot, one can observe the 
existence of quasi-steady state plateau after short time. Notice that the magnitude of $\hat{C}^2$ 
in the steady state is larger than the value expected from thermalized states at the 
initial temperature. The subsequent slow decrease of 
the interference contrast is due to the effect of temperature in the initial state coming 
from the small difference of the two quasi-condensates. 
Such development of the plateau at larger value of interference contrast than 
the one for equilibrium state of the initial temperature indicates the phenomenon 
of prethermalization.

\section{Conclusion}\label{section:conclusion}
In this work, we have shown how noise captured by full distribution functions 
can be used to study the dynamics of many-body system in one dimension. 
The analytical results of joint distribution functions obtained in Sec~\ref{section:nomixing} allow not only 
the simple understanding of 
distribution functions from spin-wave picture, but also an intuitive visualization of the correlation
in one dimensional system. Using this picture, we have also shown that the phenomena of prethermalization occur in 
one dimensional dynamics. The thermal-like behaviors of the prethermalized state is revealed through the full distribution functions that contain information about the correlation functions of the arbitrary order. 
For the experimental demonstration of such prethermalizations, see Gring {\em et al}\cite{Gring2011}. 

The approach developed in this paper can be extended to other types of 
dynamics. While we focused on Ramsey type dynamics or dynamics of interference patterns for a split quasi-condensate,  
we can also change different physical parameters to induce the dynamics. 
For example, it is straightforward to apply our study to the sudden change (quench) of interaction strength\cite{Iucci2010, Lucci2009}.

In this paper, we focused on distribution functions obtained from 
Tomonaga-Luttinger Hamiltonian (\ref{hamiltonian}). It is of interest 
to extend our analysis to higher spins\cite{Barnett2009}, and 
analyze them, for example, in the presence of magnetic field\cite{Vengalattore2008}. 
Since there are more degrees of freedom in 
these systems, distributions might capture the tendency towards various 
phases such as ferromagnetic ordering. 
These questions will be analyzed in the future works. 

We thank Igor Mazets, Alexei Gorshkov and Susanne Pielawa for useful discussions. 
The authors acknowledge support from a grant from the
Army Research Office with funding from the DARPA OLE
program, Harvard-MIT CUA, NSF Grant No. DMR-07-05472,
AFOSR Quantum Simulation MURI,
the ARO-MURI on Atomtronics.
A.I. acknowledges support from the Texas Norman Hackerman Advanced
Research Program under Grant No. 01889, the Alfred
P. Sloan Foundation under Grant No. BR-5123, and J.S. is supported by 
Austrian FWF through the Wittgenstein prize.

%

\appendix

 \begin{center}
    {\bf APPENDICES}
  \end{center}
\section{Distribution function of the $z$ component of spin}
In this appendix, we calculate the distribution function of $\hat{S}^{z}_{l}$ 
in the absence of the coupling between charge and spin. 
The extension to the case in which the charge
and spin degrees of freedom mix is straightforward. 

It is convenient to evaluate the generating function 
$\braket{e^{\lambda \hat{S}^{z}_{l}}}$, 
instead of distribution function $P^{z}_{l}(\alpha)$. 
They are related by 
\begin{equation} \label{generating}
\braket{e^{\lambda \hat{S}^{z}_{l}}} = 
\int^{\infty}_{-\infty} e^{\lambda \alpha} P^{z}_{l}(\alpha) d\alpha.
\end{equation}
This equality can be checked by differentiating both sides by $\lambda$
and evaluating them at $\lambda=0$. This reproduces 
the implicit definition of $P^{z}_{l}$ in Eq. (\ref{moments}).  

Analogous to the calculation of $m$th moment of $\hat{S}^{\theta}_{l}$, 
we first express $\hat{S}^{z}_{l}$ in terms of $\gamma_{s,k}$ operators defined 
in Eq. (\ref{gamma})
\begin{flalign*}
\hat{S}^{z}_{l}(r)   =& 
\int^{l/2}_{-l/2} dr \left( \sum_{k \neq 0} (d_{s,k} \gamma_{s,k}^{\dagger} 
+ d_{s,k}^{*} \gamma_{s,-k}) e^{ikr} +\frac{n_{s,0}}{\sqrt{L}} \right), \\
d_{s,k} = & \sqrt{\frac{|k|K_{s}}{2\pi L}}
\frac{e^{ic_{s}|k|t}+2W_{k}e^{-ic_{s}|k|t}}{\sqrt{1-4|W_{k}|^2}}.
\end{flalign*}

Then, we can apply the trick introduced in Section \ref{section:nomixing}
to obtain 
\begin{eqnarray}
\braket{e^{\lambda \hat{S}^{z}_{l}}} & = & 
e^{ \lambda^2 \int^{l/2}_{-l/2} dr_{1} dr_{2} \left( \sum_{k \neq 0} |d_{s,k}|^2 e^{ik(r_{1}-r_{2})} + \frac{\braket{n^2_{s,0}}}{\sqrt{L}} \right)}.\nonumber \\
& = & 
\exp\left\{ \lambda^2 \left(
\frac{\rho \eta l^2}{4L} + \sum_{k \neq 0} 
\frac{4 |d_{s,k}|^2}{k^2} \sin^2(lk/2) \right)  \right\} 
\nonumber \label{zresult}
\end{eqnarray}
Then the following expression gives the distribution of $\hat{S}^{z}_{l}$
\begin{equation}
P^{z}_{l}(\alpha) = \frac{1}{2\sqrt{\pi\braket{ \left( \hat{S}^{z}_{l} \right)^2}}}
\exp\left( -\frac{\alpha^2}{4 \braket{ \left( \hat{S}^{z}_{l} \right)^2}} \right),
\label{zdist}
\end{equation}
where
\begin{equation}
\braket{ \left( \hat{S}^{z}_{l} \right)^2} = 
\frac{\rho \eta l^2}{4L} + \sum_{k \neq 0} \frac{4 |d_{s,k}|^2}{k^2}
\sin^2(lk/2). \nonumber
\end{equation}

\section{Expression for $C_{a,k}$ in the presence of mixing between charge and spin} \label{appendix:c}
In this section, we derive the expression of $C_{a,k}, a=s,c$ in 
Eq. (\ref{cc}). We first find the transformation from
$b_{s,k}, b_{c \uparrow, k}$ to $b_{1,k},b_{2,k}$.
Then we relate $ b_{1,k},b_{2,k}$ and $\gamma_{s,k}(t), \gamma_{c,k}(t) $. 
Combining these two transformations, we obtain $b_{s,k}$ in terms of 
$\gamma_{s,k}(t), \gamma_{c,k}(t) $, leading to the expression of 
$\hat{\phi}_{s,k}$ in terms of $\gamma_{s,k}(t), \gamma_{c,k}(t) $. 

From the relations,
\begin{eqnarray*}
\phi_{i,k} & = & -i \sqrt{ \frac{\pi}{2|k|K_{i}}}
(b_{i,k}^{\dagger} - b_{i,-k}), \\
n_{i,k} & = & \sqrt{ \frac{|k|K_{i}}{2\pi }}
(b_{i,k}^{\dagger} + b_{i,-k}), \\
b^{\dagger}_{i,k} & = & 
i \phi_{i,k} \sqrt{\frac{|k|K_{i}}{2\pi}} + n_{i,k} \sqrt{\frac{\pi}{2|k|K_{i}}},
\end{eqnarray*}
along with Eq. (\ref{diagonalize}), 
it is straightforward to obtain
\begin{flalign*}
& \left( \begin{array}{c} b^{\dagger}_{c \uparrow,-k} \\ b_{c \uparrow,k} \\
 b^{\dagger}_{s,-k} \\ b_{s,k} \end{array}\right) 
  =  
D \left( \begin{array}{c} b^{\dagger}_{1,-k} \\ b_{1,k} \\
b^{\dagger}_{2,-k} \\ b_{2,k} \end{array}\right),
\end{flalign*}
where 
\begin{widetext}
\begin{flalign}
&D  =  \frac{1}{2} \left( \begin{smallmatrix} 
\cos\kappa \left(\sqrt{\frac{\tilde{K}_{c\uparrow}}{K_{1}}}+\sqrt{\frac{K_{1}}{\tilde{K}_{c\uparrow}}}\right) &
\cos\kappa \left(-\sqrt{\frac{\tilde{K}_{c\uparrow}}{K_{1}}}+\sqrt{\frac{K_{1}}{\tilde{K}_{c\uparrow}}}\right) &
-\sin\kappa \left(\sqrt{\frac{\tilde{K}_{c\uparrow}}{K_{2}}}+\sqrt{\frac{K_{2}}{\tilde{K}_{c\uparrow}}}\right) &
-\sin\kappa \left(-\sqrt{\frac{\tilde{K}_{c\uparrow}}{K_{2}}}+\sqrt{\frac{K_{2}}{\tilde{K}_{c\uparrow}}}\right) \\
\cos\kappa \left(-\sqrt{\frac{\tilde{K}_{c\uparrow}}{K_{1}}}+\sqrt{\frac{K_{1}}{\tilde{K}_{c\uparrow}}}\right) &
\cos\kappa \left(\sqrt{\frac{\tilde{K}_{c\uparrow}}{K_{1}}}+\sqrt{\frac{K_{1}}{\tilde{K}_{c\uparrow}}}\right) &
-\sin\kappa \left(-\sqrt{\frac{\tilde{K}_{c\uparrow}}{K_{2}}}+\sqrt{\frac{K_{2}}{\tilde{K}_{c\uparrow}}}\right) &
-\sin\kappa \left(\sqrt{\frac{\tilde{K}_{c\uparrow}}{K_{2}}}+\sqrt{\frac{K_{2}}{\tilde{K}_{c\uparrow}}}\right) \\ 
\sin\kappa \left(\sqrt{\frac{K_{s}}{K_{1}}}+\sqrt{\frac{K_{1}}{K_{s}}}\right) &
\sin\kappa \left(-\sqrt{\frac{K_{s}}{K_{1}}}+\sqrt{\frac{K_{1}}{K_{s}}}\right) &
\cos\kappa \left(\sqrt{\frac{K_{s}}{K_{2}}}+\sqrt{\frac{K_{2}}{K_{s}}}\right) &
\cos\kappa \left(-\sqrt{\frac{K_{s}}{K_{2}}}+\sqrt{\frac{K_{2}}{K_{s}}}\right) \\
\sin\kappa \left(-\sqrt{\frac{K_{s}}{K_{1}}}+\sqrt{\frac{K_{1}}{K_{s}}}\right) &
\sin\kappa \left(\sqrt{\frac{K_{s}}{K_{1}}}+\sqrt{\frac{K_{1}}{K_{s}}}\right) &
\cos\kappa \left(-\sqrt{\frac{K_{s}}{K_{2}}}+\sqrt{\frac{K_{2}}{K_{s}}}\right) &
\cos\kappa \left(\sqrt{\frac{K_{s}}{K_{2}}}+\sqrt{\frac{K_{2}}{K_{s}}}\right) \\
\end{smallmatrix}\right),  \label{dmatrix}
\end{flalign}
\end{widetext}
where $\tilde{K}_{c\uparrow} = K_{c\uparrow}/\sqrt{s_{c}}$.

Next, we relate $\gamma_{a,k}(t), a=c,s$ operators to $b_{1},b_{2}$. 
At $t=0$, we have the relation between $\gamma_{a,k}(0), a=c,s$
and $b_{c \uparrow, k}$ and $b_{s, k}$ as described in Sec.~\ref{section:mix}. 
Since operators $b_{c \uparrow, k}$ and $b_{s, k}$ are related to 
$b_{1, k}$ and $b_{2, k}$ through the matrix $D$ in Eq.(\ref{dmatrix}),
we can express $\gamma_{a,k}(0)$ as a linear combinations of 
$b_{1, k}$ and $b_{2, k}$. The time evolution of  $\gamma_{a,k}(0)$
is quite simple now because Hamiltonians are diagonal in the basis
$b_{1, k}$ and $b_{2, k}$. These considerations lead to
the relations
\begin{flalign*}
&\left( \begin{array}{c} \gamma^{\dagger}_{c,-k}(t) \\ \gamma_{c,k}(t) \\
 \gamma^{\dagger}_{s,-k}(t) \\ \gamma_{s,k}(t) \end{array}\right) 
  =  
E_{k} \left( \begin{array}{c}e^{-ic_{1}|k|t} b^{\dagger}_{1,-k}  \\e^{ic_{1}|k|t} b_{1,k}  \\
e^{-ic_{2}|k|t}b^{\dagger}_{2,-k}  \\ e^{ic_{2}|k|t} b_{2,k}  \end{array}\right), \\
&E_{k}  = F_{k} D, \\  
&F_{k} =  \left(\begin{smallmatrix}
1&
0 &
0 &
0  \\ 
0 &
1 &
0 &
0 \\
 0&
 0&
 \frac{1}{\sqrt{1-4|W_{k}|^2}} & 
 \frac{-2W_{k}}{\sqrt{1-4|W_{k}|^2}} &  \\ 
  0&
 0&
  \frac{-2W_{k}}{\sqrt{1-4|W_{k}|^2}} & 
  \frac{1}{\sqrt{1-4|W_{k}|^2}}
  \\
\end{smallmatrix}\right).
\end{flalign*}

Now define a matrix $G(k) = DE_{k}^{-1}$ so that 
\begin{displaymath}  
\left( \begin{array}{c} b^{\dagger}_{c,-k} \\ b_{c,k} \\
b^{\dagger}_{s,-k} \\ b_{s,k} \end{array}\right)  = G(k)
\left( \begin{array}{c} \gamma^{\dagger}_{c,-k} \\ \gamma_{c,k} \\
 \gamma^{\dagger}_{s,-k} \\ \gamma_{s,k} \end{array}\right)
 \end{displaymath}
Then finally $C_{i,k}$ are given by
\begin{eqnarray}
C_{s,k} & = & -i \sqrt{\frac{\pi}{2L|k|K_{s}}}(G_{34}(k) - G_{44}(k)), \nonumber \\
C_{c,k} & = & -i \sqrt{\frac{\pi}{2L|k|K_{s}}}(G_{32}(k) - G_{42}(k)), \label{expressionofcs}
\end{eqnarray}
where $G_{ij}(k)$ are the matrix elements of $G(k)$.

\section{$k=0$ contribution in the presence of mixing between charge and spin} \label{appendix:k0}
In this appendix, we evaluate 
Tr$\left\{e^{ i(\sum_{i} s_{i})\phi_{s,0}/\sqrt{L} } \rho(t) \right\} $. 
We first obtain the operator $e^{i \left(\sum_{i}s_{i}\right) \phi_{s,0}}$ 
after time evolution as 
\begin{flalign}
&e^{iHt} e^{i \left(\sum_{i}s_{i}\right) \phi_{s,0}} e^{-iHt}  =  \nonumber \\
&\exp{\left(i\frac{ \sum_{i}s_{i}}{\sqrt{L}} (A \phi_{s,0} + 
A' n_{s,0} + \frac{B}{\sqrt{s_{c}}} \phi_{c,0} + \frac{B'}{\sqrt{s_{c}}} n_{c,0})\right)}.
\label{k0operator}
\end{flalign}
Coefficients $A,A',B,B'$ can be found as follows.
$k=0$ part of the Hamiltonian is given by  
$H_{0} = \frac{\pi c_{1}}{2K_{1}} n_{1,0}^2 + \frac{\pi c_{2}}{2K_{2}} n_{2,0}^2$
(see Eq.(\ref{spindiagonalhamiltonian})). 
Using the commutation relation
$[n_{i,0}, \phi_{i,0}] = -i$, we have $ e^{iHt} \phi_{i,0} e^{-iHt}
= \phi_{i,0} + \frac{\pi c_{i}}{K_{i}} n_{i,0}t$. 
With the relation,
$\phi_{s,0} = \sin\kappa \phi_{1,0} + \cos\kappa \phi_{2,0}$,
we obtain 
\begin{flalign*}
&e^{iHt} e^{\left(\sum_{i}s_{i}\right) \phi_{s,0}} e^{-iHt} =  \\
&
e^{i\frac{ \sum_{i}s_{i}}{\sqrt{L}}
\left\{ \sin\kappa  (\phi_{1,0} + \frac{\pi c_{1}}{K_{1}} n_{1,0}t) 
+ \cos\kappa (\phi_{2,0} + \frac{\pi c_{2}}{K_{2}} n_{2,0}t) \right\}
 }.
\end{flalign*}
Now by transforming back to $c,s$ basis through Eq.(\ref{diagonalize}), we find 
\begin{eqnarray*}
A & = & 1, \\
A' & = & \sin^2\kappa \frac{\pi c_{1}}{K_{1}}t + 
\cos^2\kappa  \frac{\pi c_{2}}{K_{2}} t,\\
B & = & 0,\\
B' & = & \sin\kappa \cos\kappa  
\left( \frac{\pi c_{1}}{K_{1}} - \frac{\pi c_{2}}{K_{2}} \right)t.
\end{eqnarray*}
Now that we know the operator after time-evolution Eq. (\ref{k0operator}), 
we evaluate it in the initial state.

Initial state of the spin sector is $\ket{\psi_{s,k=0}}$ 
in Eq. (\ref{spininitial_state}). Since this state is Gaussian,
we have the simple result as follows,
\begin{eqnarray*}
& &\bra{\psi_{s,k=0}}\exp\left( \left(\sum_{i}s_{i}\right)
(A \phi_{s,0} + A' n_{s,0})/\sqrt{L}  \right) \ket{\psi_{s,k=0}}. \\
&  & =
\exp\left( -\left(\sum_{i}s_{i}\right)^2 
\left(\frac{1}{4\rho\eta L} + (A')^2 \frac{\rho\eta}{4L}\right) \right)
\end{eqnarray*}
For charge sector, $k=0$ part of the initial density matrix is 
$\frac{1}{\mathcal{N}_{c0}}\exp\left(-\beta \frac{\pi c_{c\uparrow}}{2K_{c\uparrow}} n_{c,0}^2 \right)$,
where $\mathcal{N}_{c0}$ is the normalization 
$\mathcal{N}_{c0}$ =Tr$\left(\exp\left(-\beta \frac{\pi c_{c\uparrow}}{2K_{c\uparrow}} n_{c,0}^2 \right)\right)$.
The evaluation of the charge sector yields
\begin{flalign*}
&\frac{1}{\mathcal{N}_{c0}} \textrm{Tr} \left\{\exp\left( i\left(\sum_{i}s_{i}\right) B' \frac{n_{c,0}}{\sqrt{s_{c} L}} \right)
\exp\left(-\beta \frac{\pi c_{c\uparrow}}{2K_{c\uparrow}} n_{c,0}^2 \right)\right\}\\
&  =  \exp\left( -\left(\sum_{i}s_{i}\right)^2  (B')^2 
\frac{\tilde{K}_{c\uparrow}}{2\pi c_{c\uparrow} \beta L} \right).
\end{flalign*}
Collecting the results above, we conclude
\begin{flalign*}
&\braket{e^{ i(\sum_{i} s_{i})\phi_{s,0}/\sqrt{L} }} \nonumber \\
&= \exp\left( -\frac{\left(\sum_{i}s_{i}\right)^2 }{4L}
\left\{ \frac{1}{\rho\eta } + \rho\eta(A')^2 
+(B')^2 \frac{2\tilde{K}_{c\uparrow}}{\pi c_{c\uparrow} \beta }\right\} \right). \nonumber
\end{flalign*}

\end{document}